\magnification=1200
\baselineskip=14pt
\overfullrule=0pt

\def\M{{\cal M}}
\def\d{{\rm d}}
\def\G{\gamma}
\def\p{\partial}
\centerline{\bf ON THE INTEGRABLE GEOMETRY OF
SOLITON EQUATIONS}
\medskip
\centerline{{\bf AND N=2 SUPERSYMMETRIC GAUGE THEORIES}\footnote*
{Work supported in part by the National Science Foundation
under grant DMS-95-05399}}
\bigskip
\centerline{I. M. Krichever$^{\dagger}$ and  D.H. Phong$^{\ddagger}$}
\bigskip
\centerline{$\dagger$ Landau Institute for Theoretical Physics}
\centerline{Moscow 117940, Russia}
\centerline{and}
\centerline{Department of Mathematics}
\centerline{Columbia University}
\centerline{New York, NY 10027}
\centerline{e-mail: krichev@math.columbia.edu}
\bigskip
\centerline{$\ddagger$ Department of Mathematics}
\centerline{Columbia University}
\centerline{New York, NY 10027}
\centerline{e-mail: phong@math.columbia.edu}
\bigskip
\centerline{\bf Abstract}
\bigskip
We provide a unified construction of the symplectic forms
which arise in the solution of both N=2 supersymmetric
Yang-Mills theories and soliton equations. Their phase spaces
are Jacobian-type bundles over the leaves of
a foliation in a universal configuration space. On one hand,
imbedded
into finite-gap solutions of soliton equations, these
symplectic forms assume explicit
expressions in terms of the auxiliary Lax pair, expressions which
generalize the well-known Gardner-Faddeev-Zakharov
bracket for KdV to a vast class of 2D integrable models; on the other hand,
they determine completely the effective Lagrangian and BPS spectrum
when the leaves are identified with the moduli space
of vacua of an N=2 supersymmetric gauge theory. For SU($N_c$) with $N_f\leq N_c+1$
flavors, the spectral curves we obtain this way agree
with the ones derived by Hanany and Oz and others from
physical considerations.
 
\vfill\break
\centerline{\bf I. INTRODUCTION}
\bigskip
A particularly striking aspect of
the recent solutions of N=2 supersymmetric Yang-Mills
theories [1-4] has been the emergence of
integrable structures [5-6], structures which had surfaced
in the completely different context of
soliton equations and their Whitham-averaged
counterparts [7-9]. On the gauge theory side,
the moduli space of inequivalent vacua is
identified with a moduli space of certain compact
Riemann surfaces $\Gamma$, and both the effective
Lagrangian and the Bogomolny-Prasad-Sommerfeld spectrum
can be read off from the periods of a {\it meromorphic} 1-form
$\d\lambda$ on $\Gamma$.
A defining property of $\d\lambda$ is that its external derivative 
$\delta\d\lambda$ be
a {\it holomorphic} symplectic form $\omega$ on the total space of the bundle
whose fiber is the Jacobian $J(\Gamma)$ of $\Gamma$.
On the soliton side, $\Gamma$ coincides with the spectral curve of
certain Toda or spin chain equations [5][10], and $\d\lambda$
with the pre-potential of their Whitham equations [9]. Moreover,
the symplectic form $\omega$ coincides with the symplectic structure
of these chains, considered as finite-dimensional Hamiltonian
systems.
\medskip
The most general type of finite-dimensional integrable system
in soliton theory is a space of algebraic-geometric,
or {\it finite-gap}, solutions to a soliton equation [11]. In their work in the
early 1980's, 
Novikov and Veselov [12] had proposed a 
notion of analytic symplectic form and a Hamiltonian theory for 
algebraic-geometric
solutions connected with {\it hyperelliptic curves}.
Most of the symplectic
forms $\omega$ which have arisen so far in supersymmetric gauge theories are
special cases of the Novikov-Veselov forms. As shown in the original 1994
Seiberg-Witten [1] work on SU(2) gauge theories with hypermultiplets in the 
fundamental
representation,
and in the subsequent Donagi-Witten work [6] on the SU(2) theory
with matter in the adjoint representation, the symplectic forms
are also the true indicators of integrability, and often suffice
to identify the spectral curves themselves.    
\medskip
The main goal of this paper is to build more systematically
the common foundations of the above two theories.
We construct general algebraic-geometric symplectic forms
defined on phase spaces ${\cal N}^g$ which are Jacobian-type bundles 
over leaves ${\cal M}$ of
a foliation on a universal configuration space. This last space
is a fundamental ingredient of our approach, and is defined itself
as the moduli space of {\it all} algebraic
curves with fixed jets of local coordinates at a fixed number
of punctures (c.f. Section II).

Our construction of symplectic forms and their phase
spaces is entirely geometric. To make contact with
integrable models, the phase space ${\cal N}^g$ for
each leaf is imbedded in a corresponding space
of functions as a moduli space of algebraic-geometric solutions to the
soliton equation. The algebraic-geometric symplectic form $\omega$ is
identified with the restriction to ${\cal N}^g$ of a symplectic form 
$\widehat
\omega$ defined on the space of functions. Remarkably,
in terms of the auxiliary Lax pair 
of the soliton equation, the symplectic forms 
$\widehat\omega$ assume completely explicit
and simple expressions.

As an example, consider a two-dimensional soliton equation with the flat
curvature
representation
$$
[\p_y-L, \p_t -A]=0\leftrightarrow L_y-A_y+[L,A]=0, 
$$
where
$$
L=\sum_{i=1}^n u_i(x,y,t)\p_x^i, \ A=\sum_{i=1}^m v_i(x,y,t)\p_x^i
$$
are linear operators with $(N\times N)$ matrix coefficients. Then the
symplectic form $\omega$ is given by (c.f. Section IV)
$$
\omega=-\sum_{\alpha=1}^N{\rm Res}_{P_{\alpha}}
{<\delta\psi^+\wedge(L^{(1)}\delta A-A^{(1)}\delta L)\psi>
\over<\psi^+\psi>} \d p\eqno(1)
$$
where $P_{\alpha}$ are punctures on the surface,
$\psi,\psi^+$ are the Baker-Akhiezer and dual Baker-Akhiezer
functions, $L^{(1)}$, $A^{(1)}$ are
the first descendants of the operators $L$ and $A$
$$
L^{(1)}=\sum_{i=1}^n iu_i(x,y,t)\p_x^{i-1}, \
A^{(1)}=\sum_{i=1}^m iv_i(x,y,t)\p_x^{i-1} 
$$
and $\d p$ is the differential of the quasi-momentum for the Baker-Akhiezer
function.
The formula (1) can be rewitten in turn in terms of the $\delta u_i$'s.
In this form, they unify many of
the known symplectic forms for spatially 1D models, including e.g.
the Gardner-Faddeev-Zakharov form $<\delta u\wedge\int^x\delta u>$
familiar from KdV. More importantly, they also provide 
seemingly new symplectic structures
for many 1D models, as well as a vast generalization
to a whole variety of 2D integrable models for which no symplectic
form had been available so far.
\medskip

There are strong indications that the above universal configuration space can
also serve
as a universal space for the effective Lagrangians
of $N=2$ supersymmetric gauge theories. Indeed, it suffices
to imbed the moduli space of vacua of the gauge theory as a leaf
in the universal configuration space, in order to obtain the desired
Seiberg-Witten fibration as a pull-back of the existing fibration.
Geometric considerations suggest a natural imbedding
in the case of SU($N_c$) Yang-Mills theories with $N_f\leq 2N_c$
flavors. It is intriguing that this imbedding reproduces the known solutions
of [1-3] when $N_f\leq N_c+1$, but diverges from the
formulas conjectured by Hanany
and Oz [3] for $N_c+2\leq N_f\leq 2N_c$ .
A central issue here is a proper identification
of coordinates for the universal configuration space with
the order parameters of the Yang-Mills theories. This can only be
resolved by a more careful analysis of the monodromies
of the resulting effective action at weak coupling. We
shall return to this issue elsewhere.

\vfill\break

\centerline{\bf II. THE UNIVERSAL CONFIGURATION SPACE}
\bigskip
We shall construct the universal configuration space as a moduli space of
Riemann surfaces $\Gamma$ with $N$ punctures
$(P_{\alpha})_{\alpha=1}^N$, and two Abelian integrals $E$
and $Q$ with poles of orders at
most $n=(n_{\alpha})_{\alpha=1}^N$, $m=(m_{\alpha})_{\alpha=1}^N$
at the punctures. In view of the subtleties inherent to the multiple-valuedness
of Abelian integrals, it is convenient to proceed as follows.

We define an $n_{\alpha}$-jet 
$[z_{\alpha}]_{n_{\alpha}}$
of coordinates near a puncture
$P_{\alpha}$ to be an equivalence class of coordinates $z_{\alpha}$ near 
$P_{\alpha}$, with
$z_{\alpha}$ and $z_{\alpha}'$ equivalent if
$z_{\alpha}'=z_{\alpha}+O(z_{\alpha}^{n_{\alpha}+1})$. Evidently,
the space of $n_{\alpha}$-jets of coordinates near $P_{\alpha}$
has finite dimension, equal in fact to $n_{\alpha}$. Henceforth,
we let $P_1$ be a marked puncture. In presence of a jet
$[z]_n$ near $P_1$, we can define an Abelian integral
$Q$ as a pair $(\d Q,c_Q)$, where $\d Q$ is a meromorphic differential
(or {\it Abelian differential} ) on the surface $\Gamma$,
and
$$Q=\sum_{k=-m}^{\infty}c_kz^{k}+c_Q+R^Q{\rm log}\,z\eqno(2)
$$
if $\d Q=\d(\sum_{k=-m}^{\infty}c_kz^k)+R^Q{dz\over z}$
is the expansion of $\d Q$ near $P_1$. By integrating
$\d Q$ along paths, we can then extend the Abelian integral
$E$ holomorphically to
a neighborhood of any point in $\Gamma\setminus\{P_1,\cdots,P_N\}$. The
analytic continuation will depend in general on the path, and we also 
keep track of its homotopy class.   

Fix now the multi-indices $n=(n_1,\cdots,n_N)$ and $m=(m_1,\cdots,m_N)$. The
universal configuration space $\M_g(n,m)$ can then be defined as
$$\M_g(n,m)=\{\Gamma,P_{\alpha},[z_{\alpha}]_{n_{\alpha}};E, Q\}
\eqno(3)
$$  
where $\Gamma$ is a smooth genus $g$ Riemann surface with $N$
ordered points $P_{\alpha}$, $[z_{\alpha}]_{n_{\alpha}}$ is an
$n_{\alpha}$-jet of coordinates near each $P_{\alpha}$,
and 
$E$, $ Q$ are Abelian integrals with the following expansions
near the punctures $P_{\alpha}$
$$
\eqalignno
{E&=z_1^{-n_1}+c_E+R^E_1{\rm log}\, z_1+O(z_1)\cr
\d E&=\d(z^{-n_{\alpha}}+O(z_{\alpha}))+R_{\alpha}^E
{\d z_{\alpha}\over z_{\alpha}}\cr
Q&=\sum_{k=1}^{m_1}c_{1,k}z_1^{-k}+c_Q+R_1^Q{\rm log}\,z_1+O(z_{\alpha})\cr
\d
Q&=\d(\sum_{k=1}^{m_{\alpha}}c_{\alpha,k}z_{\alpha}^{-k}+O(z_{\alpha}))+R_{\alpha}^Q
{\d z_{\alpha}\over z_{\alpha}}
&(4)\cr}
$$
The space $\M_g(n,m)$ is a complex manifold
with only orbifold singularities.
Its complex dimension is equal to
$$
{\rm dim}\M_g(n,m)
=5g-3+3N+\sum_{\alpha=1}^N(n_{\alpha}+m_{\alpha})
\eqno(5)
$$
for $g>1$. Indeed, $3g-3+N$ parameters account for the moduli space of genus $g$
Riemann surfaces with $N$ punctures, $\sum_{\alpha=1}^Nn_{\alpha}$ parameters
for the jets of coordinates $[z_{\alpha}]_{n_{\alpha}}$, $N+g$ 
parameters for the Abelian integral $E$ ($N-1$ parameters for the
independent residues
since their sum is 0, 1 parameter for the constant $c_E$,
and $g$ parameters for the Abelian differentials which can
be added without modifying the singular
expansions of (4)), and finally $N+\sum_{\alpha=1}^Nm_{\alpha}$
for the Abelian integral $Q$. Alternatively, the number of degrees of freedom
of an Abelian integral $E$ with poles of order $n=(n_{\alpha})$
is $1+\sum_{\alpha=1}^N(n_{\alpha}+1)-1+g=N+g+\sum_{\alpha=1}^Nn_{\alpha}$,
where the first 1 corresponds to the additive constant, and the remaining
integer on the left is the dimension of meromorphic
differentials with poles of order $\leq n_{\alpha}+1$ at each $P_{\alpha}$.
Together with a similar counting for $Q$ and the dimension of the moduli 
space of Riemann surfaces
with punctures, we recover (4). Note that (4) gives the right
dimension for $g=0,1; N>0$, although the counting of the
dimension is slightly different.

\bigskip
We can introduce explicit local coordinates on $\M_g(n,m)$ (for their
re\-la\-tion to the Whi\-tham theory, see in [9]).
The first of these consist of the residues of the differentials
$\d E$ and $\d Q$
$$
R_{\alpha}^E={\rm Res}_{P_{\alpha}}\d E,\ 
R_{\alpha}^Q={\rm Res}_{P_{\alpha}}\d Q,\ \alpha=2,\cdots,N\eqno(6)
$$
The next set of coordinates is only local
on the universal configuration space, and requires some choices.
First, we cut apart the Riemann surface $\Gamma$
along a homology basis
$A_i,B_j$, $i,j=1,\cdots,g$, with the canonical intersection
matrix $A_i\cdot A_j=B_i\cdot B_j=0, A_i\cdot B_j=\delta_{ij}$.
By selecting cuts from $P_1$ to $P_{\alpha}$ for each $2\leq\alpha\leq N$,
we obtain a well-defined branch of the Abelian integrals
$E$ and $Q$. Locally on the universal configuration space,
this construction can be carried out continuously,
with paths homotopic by deformations not crossing
any of the poles.  
We consider first the case when $n_{\alpha}$ is at least 1.
Then there exists a unique local coordinate $z_{\alpha}$ in 
the jet $[z_{\alpha}]_{n_{\alpha}}$ such that
$$
E=z_{\alpha}^{-n_{\alpha}}+R_{\alpha}^E{\rm log}\,z_{\alpha}
$$
in the neighborhood of $P_{\alpha}$. We set then
$$
\eqalignno{
T_{\alpha,k}&={1\over k}{\rm Res}_{P_{\alpha}}(z_{\alpha}^kQ\d E),\ \alpha=1,
\ \cdots,N,\ k=1,\cdots,
n_{\alpha}+m_{\alpha},\cr
T_{\alpha,0}&={\rm Res}_{P_{\alpha}}(Q\d E),\ \alpha=2,\cdots,N&(7)\cr}
$$
When $n_{\alpha}=0$, we choose the coordinate $z_{\alpha}$
by demanding that $E=R_{\alpha}^E{\rm log}\,z_{\alpha}$. 

A final set
of coordinates for the universal configuration space can now be defined by
$$
\eqalignno{
\tau_{A_i,E}&=\oint_{A_i}\d E,\ \tau_{B_i,E}=\oint_{B_i}\d E&(8)\cr  
\tau_{A_i,Q}&=\oint_{A_i}\d Q,\ \tau_{B_i,Q}=\oint_{B_i}\d Q&(9)\cr  
a_i&=\oint_{A_i}Q\d E,\ i=1,\cdots,g&(10)\cr}
$$
Let ${\cal D}$ be the open set in $\M_g(m,n)$ where
the zero divisors of $\d E$ and $\d Q$, namely the sets
$\{\G;\d E(\G)=0\}$ and $\{\G;\d Q(\G)=0\}$, do not intersect.
Then
\medskip
\noindent{\bf Theorem 1}. {\it {\rm (a)} Near each point in ${\cal D}$, 
the $5g-3+3N+\sum_{\alpha=1}^N(n_{\alpha}+m_{\alpha})$ functions
$R_{\alpha}^E$, $R_{\alpha}^Q$, $T_{\alpha,k}$,
$\tau_{A_i,E}$, $\tau_{B_i,E}$, $\tau_{A_i,Q}$,
$\tau_{B_i,Q}$, $s_i$ of (6-10) have linearly independent
differentials, and thus define a local holomorphic
coordinate system for $\M_g(n,m)$; {\rm (b)} The joint level sets of
the functions (6-9) define a smooth
$g$-dimensional foliation of $\cal D$, independent of the choices
we made to define the coordinates themselves.}
\bigskip
To lighten the exposition of the paper, we postpone the proof of
Theorem 1 to the Appendix.
\bigskip
The universal configuration space $\M_g(n,m)$ is the base space
for a hierarchy of fibrations ${\cal N}_g^k(n,m)$ of particular interest to us.
These are the fibrations whose fiber above each point
of $\M_g(n,m)$ is the $k$-th symmetric power $S^k(\Gamma)$
of the curve $\Gamma$.  
The first, ${\cal N}_g^{k=1}(n,m)\equiv{\cal N}_g(n,m)$, is just a version of the
universal curve, where the fiber above a point
of $\M_g(n,m)$ the Riemann surface $\Gamma$ itself.
The fibration ${\cal N}_g^g(n,m)$ with $k=g$ can be
viewed as a universal Jacobian, by identifying
$S^g(\Gamma)$ with the Jacobian $J(\Gamma)$
via the Abel map
$$
(\G_1,\cdots,\G_g)\rightarrow\phi_j=\sum_{i=1}^g\int_{P_1}^{\G_i}\d\omega_j
$$
The fibrations with $k>g$ enter the study of matrix solitons,
as we shall see later.
\medskip
Consider now a leaf of the foliation described in Theorem 1,
denoted just by $\M$ for simplicity, and let ${\cal N}$
and ${\cal N}^g$ denote the above fibrations restricted to
${\cal M}$. To define a symplectic form $\omega_{\M}$ on ${\cal N}^g$,
we begin by discussing some aspects of differentials
of the Abelian integrals $E$ and $Q$ on ${\cal N}$.

The first key observation is that, 
although the Abelian differentials $E$ and $Q$ are multi-valued
functions on the universal fibration of curves
over the moduli space $\M_g(n,m)$, their differentials
are well-defined on ${\cal N}$. Indeed, by our normalizations,
$E$ and $Q$ are well-defined in a small neighborhood of the
puncture $P_1$. Their analytic continuations by different paths
can only change by multiples of their residues or periods
along closed cycles. Since along a leaf of the foliation, these
ambiguities remain constant, they disappear upon differentiation.
We shall denote these differentials along the fibrations by
$\delta E$ and $\delta Q$. Acting on vectors tangent to the fiber,
they reduce of course to the usual differentials
$\d E$ and $\d Q$. 

Next, we note that the choice of an Abelian integral, say $E$,
also provides us with a {\it meromorphic} connection $\nabla^E$ on
${\cal N}$. Indeed,
at any point of ${\cal N}$, the variety $E=constant$
is intrinsic and transversal to the fiber. This means we
can differentiate functions on ${\cal N}$ along
vector fields $X$ on $\M$, simply
by lifting these vector fields
to vector fields tangent to the varieties $E=constant$.
More generally, we can differentiate 1-forms $\d f$ on the fibration
of curves $\Gamma$ by setting
$$
\nabla_X^E(\d f)=\nabla_X^E({\d f\over\d E})\d E\eqno(11)
$$
Furthermore, the differential
$\d Q$ of any Abelian integral $Q$
along the fiber can be extended to a 1-form
on the whole manifold ${\cal N}$ by making it zero along
$E=constant$. We still denote this differential by
$\d Q$. Equivalently, we can trivialize the fibration
${\cal N}$ with the variables $a_1,\cdots,a_g$ along the leaf ${\cal M}$,
and the variable $E$ along the fiber. The form $\d Q$ we defined
previously 
coincides with ${\d Q\over\d E}\d E$, where $\d E$ is viewed as one of the 
elements
of the basis of 1-forms for ${\cal N}$ in this coordinate
system. The full differential $\delta Q$ is given by
$$
\delta Q=\d Q+\sum_{i=1}^g{\partial Q\over\partial a_i}da_i\equiv\d Q+\delta^EQ
$$  
For $E$, we have $\delta E=\d E$, and the above connection reduces to 
$\nabla_{\partial_{a_i}}^E=\partial_{a_i}$.
\medskip
If we consider now the full differential $\delta(Q\d E)$ on ${\cal N}$,
it is readily
seen that it is well-defined, despite the multivaluedness
of $Q$. In fact, the partial derivatives $\partial_{a_i}(Q\d E)$ 
along the base ${\cal M}$ are holomorphic, since the singular
parts of the differentials as well as the ambiguities are all fixed.
In particular,
$$
{\partial\over\partial s_i}(Q\d E)=\d\omega_i
$$
where $\d\omega_i$ is the basis of normalized holomorphic differentials
$$
\oint_{A_i}\d\omega_j=\delta_{ij},\ \oint_{B_i}\d\omega_j=\delta_{ij}
$$
and $\tau_{ij}$ the period matrix of $\Gamma$.
We can now define the desired symplectic form $\omega_{\cal M}$ on 
${\cal N}^g$ by
$$
\omega_{\cal M}=\delta\big(\sum_{i=1}^gQ(\G_i)\d E(\G_i)\big)=
\sum_{i=1}^g\delta Q(\G_i)\wedge\d E(\G_i)=
\sum_{i=1}^g\d s_i\wedge\d\omega_i\eqno(12)
$$
In many situations, we need to go beyond
the case ${\cal N}^g$, and consider the fibration
${\cal N}^k$ for $k>g$. The above form on
the leaves $\M$ is then degenerate. However,
a non-degenerate form for $k>g$ can be obtained by restricting
${\cal N}^k$ to the larger leaves $\tilde{\cal M}$ of the fibration
corresponding to the level sets
of all the functions (6-9) except for $T_{\alpha,0}$.
Thus we can set
$$
\omega_{\tilde{\cal M}}=
\sum_{i=1}^{k}\delta Q(\G_i)\wedge\d E(\G_i)=
\sum_{i=1}^g\d s_i\wedge\d\omega_i+\sum_{\alpha=2}^{k-g+1} \d T_{\alpha,0}\wedge
\widetilde{\d\omega_{\alpha}}
$$
where $\widetilde{\d\omega_{\alpha}}$ is a normalized differential with
only simple poles at $P_1$ and $P_{\alpha}$, and residues $-1$ and $1$
respectively. This form $\omega_{\tilde{\M}}$ is meromorphic. So far we
do not know its role,
if any,
in supersymmetric gauge theories, but it is  responsible for the
symplectic structure
of finite-gap solutions to integrable equations with matrix Lax operators
(see Section IV).

\bigskip
\centerline{\bf III. THE SYMPLECTIC FORM FOR THE KP HIERARCHY}
\bigskip
Remarkably, the natural geometric symplectic form
constructed in the last section leads directly to a
Hamiltonian structure for soliton equations. We shall derive
it explicitly in the case of the Kadomtsev-Petviashvili (or KP)
hierarchy, and show that it reduces, in the case
of KdV, to the familiar Gardner-Faddeev-Zakharov symplectic form
on the space of functions of one variable.
\bigskip
\noindent {\it 1) Solutions of the KP Hierarchy}

We begin by recalling some notions from the
algebraic-geometric theory of solitons [11]. A fundamental principle is that
to each data consisting of
a smooth Riemann surface $\Gamma$ with $N$ punctures $P_{\alpha}$,
local holomorphic coordinates $z_{\alpha}$ around each puncture,
and a divisor $(\gamma_1,\cdots,\gamma_g)$, there corresponds
a sequence 
$
\{u_{i,n}^{\alpha}(t)\}_{1\leq i\leq n-2,\ 2\leq n<\infty}
$
of complex, quasi-periodic functions of an arbitrarily large number
of variables $t=(t_{n,\alpha})_{n=1}^{\infty}$, which are solutions of
an infinite integrable
basic hierarchy of soliton equations. The $\{u_{i,n}^{\alpha}\}$ are obtained via
the Baker-Akhiezer function $\psi(t,z), \ z\in \Gamma$, which is
the unique function meromorphic away from $P_{\alpha}$,
with simple poles at $\G_i$, $i=1,\cdots,g$, and which
admits the following essential
singularity at $P_{\alpha}$
$$
\psi(t,z_{\alpha})= {\rm
exp}(\sum_{n=1}^{\infty}t_{n,\alpha}z_{\alpha}^{-n})(1+\sum_{i=1}^{\infty}
\xi_{\alpha,i}(t)z_{\alpha}^i)\eqno(13)
$$
For each $n$, it is then straightforward to derive
recursively a unique
linear differential operator in ${\partial/\partial t_{1,\alpha}}$ 
$$
L_n^{\alpha}=({\partial\over\partial t_{1,\alpha}})^n+
\sum_{i=1}^{n-2}u_{i,n}^{\alpha}({\partial\over\partial
t_{1,\alpha}})^i\eqno(14)
$$
so that
$$
({\partial\over\partial t_{n,\alpha}}-L_n^{\alpha})\psi(t,z)=0
$$
The coefficients of $L_n$ are the functions we seek, and the soliton
equations they satisfy are the partial differential equations
in the variables $t$ resulting from the commutation
relations $[{\partial\over\partial t_{n,\alpha}}-L_n^{\alpha},
{\partial\over\partial t_{m,\alpha}}-L_m^{\alpha}]=0$.
If we let
$$
\widehat{\cal N}^k=\{\Gamma,P_{\alpha},
z_{\alpha},\G_1,\cdots,\G_k\}\eqno(15)
$$ 
be the space of data of the above form, we have this way a map
from $\widehat{\cal N}^g$ to the space of sequences
of functions $\{u_{i,n}^{\alpha}(t)\}$. The space $\widehat{\cal N}^k$
is of course infinite-dimensional, but we shall soon
restrict to a more manageable finite-dimensional subspace.
The above-defined Baker-Akhiezer functions corresponds to linear operators
with scalar coefficients. In the general case, the Baker-Akhiezer 
vector-functions which are defined by (15) with $k=g+N-1$ give solutions 
to the integrable hierarchy with the flat curvature representation and 
linear operators with matrix $(N\times N)$ coefficients (see section IV).
\medskip
The type of basic hierarchy
is characterized by $N$ and $k$. For example, $N=1, k=g$ gives
the KP hierarchy, while $N=2, k=g$
gives the 2D Toda lattice hierarchy. The KdV hierarchy
is a reduction of the KP hierarchy, and in fact, all known hierarchies
are reductions of the basic ones. Henceforth, we shall
restrict our attention to the KP hierarchy, where $N=1$,
and simply drop the index $\alpha$ from our notation.
\medskip 
It is instructive to write the Baker-Akhiezer function more explicitly. 
The key ingredients are the meromorphic differentials
$\d\Omega_n$, which are characterized by their
expansion $\d\Omega_n=\d(z^{-n}+O(z))$ near $P$, and by their normalization
$$
{\rm Re}\oint_C\d\Omega_n=0\eqno(16)
$$
for any cycle $C$. The corresponding Abelian integrals $\Omega_n$
can then be defined as 
$$
\Omega_n=z^{-n}+O(z)\eqno(17)
$$
near the puncture $P$. As before, 
$\Omega_n$ can then be extended by analytic
continuation to a neighborhood of an arbitrary
point in the Riemann surface $\Gamma$. Its value
depends on the homotopy class of the path along which the
analytic continuation is performed, and we
keep track of this path as well.  The Baker-Akhiezer
function can then be expressed as [8]
$$
\psi(t,z)={\rm exp}(\sum_{n=1}^{\infty}t_n\Omega_n)\Phi(t,z)
$$
with the quasi-periodic function $\Phi(t,z)$ given by
$$
\eqalign{
\Phi(t,z)
=&
{\Theta\big(\int^z\omega_i-\sum_{j=1}^g\int^{\G_j}\omega_i
+\sum_{t=n}^{\infty}(\oint_{A_i}\d\Omega_n-\sum_{j=1}^g\tau_{ij}\oint_{B_j}
\d\Omega_n)+K\big)
\over
\Theta\big(\int^{P}\omega_i-\sum_{j=1}^g\int^{\G_j}\omega_i
+\sum_{t=n}^{\infty}(\oint_{A_i}\d\Omega_n-\sum_{j=1}^g\tau_{ij}\oint_{B_j}
\d\Omega_n)+K\big)}\cr
&\times
{\Theta\big(\int^{P}\omega_i-\sum_{j=1}^g\int^{\G_j}\omega_i+K\big)
\over
\Theta\big(\int^z\omega_i-\sum_{j=1}^g\int^{\G_j}\omega_i+K\big)}
{\rm exp}\big(-\sum_{n=1}^{\infty}t_n\sum_{j=1}^g\oint_{B_j}\d\Omega_n
\int_{P}^z\omega_j)\cr}
$$
where $\tau_{ij}$ is the period matrix of
$\Gamma$, and $K$ is the vector of Riemann constants.
We note that although the Abelian integrals $\Omega_n$ as well
as the Abel map $\int^z\omega_i$ are path dependent, this dependence
cancels in the full expression for the Baker-Akhiezer function, as it
should.
\medskip
For later use, we also recall here the main properties
of the dual Baker-Akhiezer function $\psi^{+}(t,z)$. To define it,
we note that the Riemann-Roch theorem implies that for $g$ points
$\G_1,\cdots,\G_g$ in general position, the unique meromorphic
differential 
$$
\d\Omega=\d(z^{-1}+\sum_{s=2}^{\infty}a_sz^s)\eqno(18)
$$
with double pole at $P$ and zeroes at $\G_1,\cdots,\G_g$,
must also have $g$ other zeroes. We denote these by
$\G_1^+,\cdots,\G_g^+$. The dual Baker-Akhiezer function $\psi^+(t,z)$
is then the unique function $\psi^+(t,z)$
which is meromorphic everywhere except at $P$, has at most
simple poles at $\G_1^+,\cdots,\G_g^+$, and admits the
following expansion near $P$
$$
\psi^+(t,z)
={\rm exp}(-\sum_{n=1}^{\infty}t_nz^{-n})(1+\sum_{s=1}^{\infty}\xi_s^+(t)z^s)
\eqno(19)
$$
It is then not difficult to check that the dual Baker-Akhiezer
function satisfies the equation
$$
-{\partial\over\partial t_n}\psi^+(t,z)
=\psi^+(t,z)L_n\eqno(20)
$$
where the operator $L_n$ is the same as the one of (14), and the above left
(adjoint) action of differential operators is defined by
$$
(f^+\partial_{t_1}^i)=(-1)^i(\partial_{t_1}^if^+)
$$
The coefficients $\xi_s^+$ of the expansion (19) for the dual Baker-Akhiezer
function are differential polynomials in the coefficients $\xi_s$
of the Baker-Akhiezer function. In fact, we have
$$
{\rm Res}_P\psi^+(t,z)\big(\partial_{t_1}^m\psi(t,z)\big)\d\Omega=0
$$
since the differential on the left hand side
is meromorphic everywhere, and holomorphic away from $P$. This implies that
$$
\sum_{l=0}^m\sum_{s+i+j=1}C_m^la_s\xi_i^+(\partial_{t_1}^{m-l}\xi_{j+l})=0
$$
where $a_0=1, a_1=0$, and $a_s$ are the coefficients of the
expansion (18) of $\d\Omega$ near $P$. These equations determine
$\xi_s^+$ recursively
through $\xi_s$. For example,
$$
\xi_1^+=-\xi_1,\ \xi_2^+=-\xi_2+\xi_1^2-\partial_{t_1}\xi_1
$$
\vfill\break
\noindent{\it 2) The imbedding of the leaves $\M$ into the space of KP solutions}  

Consider now the fibration ${\cal N}_g^g(n,1)$ of the last section,
in the case of a single ($N=1$) puncture $P$.
Recall that its
base $\M_g(m,1)$ consists of Riemann surfaces $\Gamma$
with a puncture $P$, of jets $[z]_{n}$
of coordinates around $P$, and of Abelian integrals $E$, $Q$ with
only poles at $P$, of order $n$ and $1$ respectively.
Its fiber over each point is simply the $g$-symmetric power
of $\Gamma$. The basic observation is that 
an element in ${\cal N}_g^g(n,1)$
gives rise to a data in $\widehat{\cal N}^g$. Indeed, 
as we already noted when introducing
local coordinates for ${\cal M}_g(n,m)$, 
the Abelian integral $E$ characterizes 
a unique holomorphic coordinate around $P$ in the given jet $[z]_n$,
satisfying  
$
E(z)=z^{-n}+R^E{\rm log}z
$
and the fact that it is in the given jet. We obtain in this way a
map 
$$(\Gamma, P, [z]_{n}, E,Q;\G_1,\cdots,\G_g)\rightarrow
(\Gamma, P, z;\G_1,\cdots\G_g)
\eqno(21)
$$ 
from ${\cal N}_g^g(n,1)$ into
the space of data $\widehat {\cal N}^g$, and hence into the space
of algebraic-geometric solutions of the KP hierarchy
$$
{\cal N}_{g,1}^g\rightarrow\{(u_{i,n}(t))_{i=1}^{n-2}\}\eqno(22)
$$
We restrict our attention now to a {\it real leaf} $\M$,
that is, a leaf of the foliation of Theorem 1,
which satisfies the following additional condition
$$
{\rm Re}\oint_C \d E={\rm Re}\oint_C\d Q=0\eqno(23)
$$
for all cycles $C$ on $\Gamma$. We say
then that the differentials
$\d E$ and $\d Q$ are {\it real-normalized}. This condition
is unambiguous, since the residues of $\d E$ and $\d Q$ are 0.
On real leaves, the Abelian differentials $\d E$ and $\d Q$
are readily recognized as the Abelian differentials
$\d Q=\d\Omega_1$ and $\d E=\d\Omega_n$ of (17),
simply by comparing their singularities at $P$. 
\medskip
Our main goal is to express the symplectic
form $\omega_{\cal M}$ we constructed earlier in terms
of forms on the space of functions $\{u_{i,n}(t)\}$. 
The functions $u_{i,n}(t)$ can be written
explicitly in terms of the coefficients $\xi_s$
of the Baker-Akhiezer function $\psi(t,z)$ with the help of the equations
$$
\sum_{i=0}^n u_{i,n}\sum_{l=0}^i C_i^l(\p_x^{i-l}\xi_{s+l})=\xi_{s+n},\ 
s=-n+2,\ldots,-1,0.
$$
They are differential polynomials in the first $n-1$ coefficients $\xi_s(t)$ 
of the expansion (13) of $\psi$.

The equations for $u_{i,n}$ are almost invertible. Let us consider them
as a system of equations for unknown functions $\xi_s, \ s=1,\ldots,n-1$ 
with $u_{i,n}$ given.
Then $\xi_s,\ s\leq n-1$ are uniquely defined by this system if we fix the 
initial data:
$$
\xi_s(t)|_{x=0}=\varphi_s(t_2,t_3,\ldots). 
$$
That ambiguity does not effect the first $n-1$ coefficients of an expansion
of the logarithmic derivative of $\psi$
\medskip
\item{} {\it The $n-1$ leading coefficients $h_1(t),\cdots,h_{n-1}(t)$
of the expansion
$$
{\partial_{t_1}\psi\over\psi}=z^{-1}+\sum_{s=1}^{\infty}h_sz^s\eqno(24)
$$
are differential polynomials
$$
h_s(t)=h_s(u_{i,n},\partial_{t_1}u_{i,n},\cdots)
$$
in the $u_{i,n}$'s. The same is true for the 
first $n-1$ coefficients $h_1^+(t),\cdots,h_{n-1}^+(t)$ of the
expansion of the dual Baker-Akhiezer function}
$$
{\partial_{t_1}\psi^+\over\psi^+}=-z^{-1}+\sum_{s=1}^{\infty}h_s^+z^s\eqno(25)
$$
\medskip
The differential polynomials $h_s$ and $h_s^+$ are universal and depend
on $n$ only. For the remainder of this section as well
as Section IV, we shall adopt a notation of greater
use in the study of soliton equations, and set
$$
x=t_1,\ p=\Omega_1$$
as the basic space-variable and its corresponding quasi-momentum.
We also fix an $n$, and set $y=t_n$.
In this notation, we have for example
$$
\eqalignno{
h_1&=-{1\over n}u_{n-2,n}\cr
h_2&={n-1\over n}u_{n-2,n;x}-{1\over n}u_{n-3,n}\cr
h_3&=-{(n-1)(n-2)\over 6n}u_{n-2,n;x}+{n+1\over 2n^2}u_{n-2,n}^2+
{n-1\over 2}u_{n-3,n;x}-{1\over n}u_{n-4,n}&(26)\cr}
$$
while the first few coefficients $h_s^+$ are given by
$$
\eqalignno{
h_1^+&={1\over n}u_{n-2,n}\cr
h_2^+&={3-n\over 2n}u_{n-2,n;x}+{1\over n}u_{n-3,n}&(27)\cr}
$$
The mean values of the polynomials $h_s(t)$ are equal
to the first coefficients of the expansion of the Abelian integral $p=\Omega_1$
$$
\eqalignno{p&=z^{-1}+\sum_{s=1}^{\infty}H_sz^s\cr
H_s&=<h_s(t)>_x\equiv{\rm lim}_{L\rightarrow\infty}\int_{-L}^L{dx\over 2L}
h_s(x,y)
&(28)\cr}
$$
The bundles ${\cal N}^g$ over real leaves ${\cal M}$ of the
foliated manifold ${\cal M}_g(n,1)$ can now be recognized
as the preimage under the map (21) of the level sets of the integrals
$H_s=<h_s>_x$.

We can now state one of our main results:
\medskip
\noindent{\bf Theorem 2}. {\it Let ${\cal N}^g$ be the Jacobian
bundle over a real leaf ${\cal M}$ of the moduli space $\M_g(n,1)$.
Then the symplectic form $\omega_{\cal N}$ of (12) can be expressed
as
$$
\omega_{\cal N}={\rm Res}_{P}{<\delta\psi^+\wedge\delta L\psi>\over
<\psi^+\psi>}\d p
=n\sum_{s=1}^{n-2}<\delta h_s\wedge\int^{t_1}\delta^* h_{n-s}^+>
\eqno(29)
$$
where $h_s=h_s(u_{i,n},\cdots)$ and $h_s^+(u_{i,n},\cdots)$ are the
above differential polynomials, the 1-forms $\delta h_s$ and $\delta^* h_s^+$
are defined respectively by
$$
\delta h_s=\sum_{i=0}^{n-2}\delta u_{i,n}\sum_{l}(-\partial_{t_1})^l
{\partial h_s\over\partial u_{i,n}^{(l)}},\   
\delta^* h_s^+=\sum_{i=0}^{n-2}\delta u_{i,n}^{(l)}\sum_{l}
{\partial h_s^+\over\partial u_{i,n}^{(l)}},
\eqno(30)
$$  
and $<\cdot>$ denotes the average with respect to both variables
$x=t_1$ and $y=t_n$ of quasi-periodic functions}
$$
<f>=lim_{L,M\rightarrow\infty}\int_{-L}^L{dx\over2 L}
\int_{-M}^M{\d y\over 2M}f(x,y)
$$
\bigskip
\noindent{\it Important Remark.} In general, finite-gap solutions to soliton 
equations
are complex meromorphic functions of all variables. This requires a more
delicate definition of averaging. Without discussing this point in detail,
we would like to mention that the averaging we adopt in this paper is 
valid at least for real and smooth
solutions of soliton equations. The corresponding constraints for 
algebraic-geometric data singles out the {\it real} part of the 
configuration space.
\medskip
\noindent{\it Proof of Theorem 2}. 
We had noted earlier that the full
differentials of two Abelian differentials $\delta p$ and $\delta E$
are well-defined on the fibration ${\cal N}$.
In terms of the imbedding (21), the differentials of $E=\Omega_n$
and $p=\Omega_1$ satisfy the following
important relation [8]
$$
\delta E=\delta p{\d E\over\d p}+
{<\psi^+\delta L_n\psi>\over<\psi^+\psi>}\eqno(31)
$$
where $\delta L_n$ is given by
$$
\delta L_n=\sum_{i=0}^{n-2}\delta u_{i,n}(t)\partial_x^i
$$
Now the symplectic form $\omega_{\cal M}$
can be expressed as
$$
\omega_{\cal M}=\sum_{s=1}^g\delta p(\gamma_s)\wedge\,\d E(\gamma_s)\eqno(32)
$$
As we had seen in the discussion leading to the equation (12), $\delta E=\d E$.
Thus we may write in view of (31)
$$
\eqalignno{\omega_{\cal N}&=
\sum_{s=1}^g\big(\d E(\G_s)-{<\psi^+\delta L_n\psi>\over
<\psi^+\psi>}(\G_s)\big){\d p\over\d E}(\G_s)\wedge\,\d E(\G_s)\cr
&=-\sum_{s=1}^g{<\psi^+\delta L\psi>\over<\psi^+\psi>}(\G_s)
\wedge\,\d p(\G_s)&(33)\cr}
$$
Our next step is to rewrite this expression
in terms of residues at the marked puncture $P$.
For this, we begin by changing coordinates on the Jacobian. Now
the Baker-Akhiezer function $\psi(t,z)$ has $g$ zeroes
$\G_s(t)$ on $\Gamma$ outside of the
puncture $P$. The Riemann-Roch Theorem implies
that these punctures coincide with
the poles $(\gamma_1,\cdots,\G_g)$ when $t=0$. Otherwise, as $t$
varies, the dependence of the zeroes
$\gamma_s(t)$ allow us to consider any set
of $g$ times $t_{i_1},\cdots,t_{i_g}$ for which
the corresponding flows are independent, as a system of coordinates
on the Jacobian $S^g(\Gamma)$. In general position,
we can choose $t_1,\cdots,t_g$ as this system. The transformation
from these coordinates to the system $(f(\G_1),\cdots,f(\G_g))$
defined by an Abelian integral $f$ on $\Gamma$
is described by the following formula
$$
\partial_{t_i}f(\G(t))={\rm Res}_{\G(t)}{\partial_{t_i}\psi(t,z)\over\psi(t,z)}\d f
\eqno(34)
$$
In the present case, we find
$$
\omega_{\cal M}=-\sum_{s=1}^g {\rm Res}_{\G_s}\left[{<\psi^+\delta L \psi>\over
<\psi^+\psi>}\wedge {\delta_t \psi\over \psi}|_{t=0}\right]\d p
\eqno(35)
$$
where 
$$
{\delta_t \psi\over \psi}=\sum_{j=1}^g {\p_j \psi(t,P)\over \psi(t,P)}\d t_j.
$$
The variation $\delta_t$ can be viewed as the restriction of the full variation 
$\delta$ to a fiber
of ${\cal N}^g$. In view of the properties of the dual
Baker-Akhiezer function, the differential $\d\Omega$ of (18)
can be recognized as
$$
\d\Omega={\d p\over <\psi^+\psi>}\eqno(36)
$$
Therefore, the differential in the right hand side of (35) is a 
meromorphic differential with poles at $P$ and the points $\G_s=\G_s(0)$
only. Thus the sum of the residues at $\G_s$
is just the opposite of the residue at $P$
$$
\omega_{\cal M}={\rm Res}_{P}\left[{<\psi^+\delta L \psi>\over
<\psi^+\psi>}\wedge {\delta_t \psi\over \psi}|_{t=0}\right]\d p
$$
Let $\delta J_s(t)$ be the coefficients of the expansion
$$
{\psi^+(\delta L \psi)\over <\psi^+\psi>} \d p=
-\sum_{s=1}^{\infty}\delta J_s(t)z^{-n-1+s}\d z\eqno(37)
$$
The equation (31) implies that
$$
<\delta J_s(t)>_x=\delta H_s=0, \ s=1,\cdots,n-1,
$$
since the mean values 
$H_s$ are fixed along the leaf ${\cal M}$ (recall $<\cdot>_x$
denotes averages with respect to $x$, keeping $t_n=y$ fixed). In particular,
$$
\omega_{\cal M}=-\sum_{j=1}^{g}<\delta J_{j+n}(t)>\wedge\, \d t_j\eqno(38)
$$
Next, the differential
$$
{\p_j\psi^+(t,P)(\delta L \psi(t,P)) \over<\psi^+\psi>} \d p 
$$
is holomorphic on $\Gamma$ except at $P$. Therefore, its residue at this
point is equal to zero and we obtain
$$
\delta J_{j+n}(t)=\sum_{s=1}^{n-1}\delta J_s(t)h^+_{n-s,i}(t)\eqno(39) 
$$
with $h_{s,j}(t)$ the coefficients of the expansion
$$
{\p_j \psi^+(t,P)\over \psi^+(t,P)}=-z^{-j}+
\sum_{s=1}^{\infty}h_{s,j}^+(t)z^s\eqno(40) 
$$
Therefore, (38) and (39) imply that 
$$
\omega_{\cal M}=-{\rm Res}_{P_0}{<\delta_t \psi^+\wedge \delta L\psi>\over
<\psi^+\psi>}\d p 
$$
We would like to replace the partial 
variation $\delta_t$ along the fiber in the
preceding formula by the full variation $\delta$.
To this effect, we show now that the residue of the contribution
of the orthogonal variation $\delta^E$, 
i.e. the variation with fixed $E(\G_s)$, vanishes 
$$
{\rm Res}_{P}{<\delta^E \psi^+\wedge \delta L\psi>\over
<\psi^+\psi>}\d p=0 
$$
Indeed, the left hand side can be rewritten as
$$
\delta^E\ \left({\rm Res}_{P_0}{\rm log}\, \psi^+{<\psi^+\delta L\psi>
\over<\psi^+\psi>}\d p\right)= 
\delta^E \left(\sum_{s}2\pi i\int_{\gamma_s^+}^{\gamma_s^+(t)}\delta^Ep\d E
\right)=0
$$
with the second equality a consequence
of the main property of differentials, $(\delta^E)^2=0$.
This establishes the first identity stated in Theorem 2.
\medskip
Next, it follows from (40) that
$$
\delta {\rm log}\, \psi^+=\delta \left(c(t_2,t_3,\ldots;k)+ 
\int_{x_0}^x \p_x \ln \psi^+\right)=
\delta \sum_{s=1}^{\infty}\left(c_s(t_2,t_3,\ldots,k)+\int_{x_0}^x h_s^+ \d x
\right). 
$$
Since $<\delta J_s>_x=0$, the above integration
constants are immaterial, and we obtain
$$
\omega_{\cal N}=-\sum_{s=1}^{n-1}
<\delta J_s\wedge \int_{x_0}^x \delta^*h_{n-s}^+\d x>.
$$
From its definition (37), $\delta J_s$ does not contain variations of
derivatives of $u_i$. Therefore, for 
$s=1,\ldots, n-1$,
$$
\delta J_s = -n {\delta h_s \over \delta u}\delta u. 
$$
Substituting in the preceding equation gives the second
identity in Theorem 2.
\bigskip
\noindent{\it 3) The symplectic form for the Korteweg-de Vries Equation}

The KdV equation
corresponds to $n=2$.
The relevant
differential operator $L$ is then the second order differential operator
$$
L=\partial_x^2+u(x,y). 
$$
and its coefficients are quasi-periodic functions $u(x,y)$ of two variables
($y=t_2$). In this case the formula (29) becomes
$$
\omega_{\cal M}=-{1\over 2}<\delta u \wedge\int_{x_0}^x \delta u\, \d x>\eqno(41) 
$$
and $\omega_{\cal M}$ is well-defined on a space of functions with 
fixed mean value in $x$
$$
<u>_x=H_1 \longrightarrow <\delta u>_x =0. 
$$
The symplectic form (41) reduces to the Gardner-Faddeev-Zakharov symplectic form
when $u(x,y)=u(x)$ is a function of one variable
only. In this case, we get a reduction of the KP equation to the KdV equation
$$
u_t={3\over 2}uu_x+{1\over 4}u_{xxx}. 
$$
Indeed, for $n=2$, the differentials
$\d p$ and $\d E$ at the puncture $P$ have the form
$$
\d p=\d(z^{-1}+O(z)),\ \ \d E=\d(z^{-2}+O(z)) 
$$
Consider the leaf of the foliation defined by these differentials
which corresponds to $\d E$ with all zero periods
$$
\oint_C \d E=0  
$$
for an arbitrary cycle $C$. In this case, the Abelian integral $E(P)$ is a 
single-valued function, with 
only a pole of second order at $P$. 
The corresponding
curve is a hyperelliptic curve given by an equation of the form 
$y^2=\prod_{i=1}^{2g+1}(E-E_i)$, and 
$P$ is the point at infinity  $E=\infty$. 
For finite-gap solutions of the KdV equation, the periods
$$
a_i =\oint_{A_i} p\d E 
$$
are canonically conjugated (with respect to the Gardner-Faddeev-Zakharov 
symplectic structure) to the angle variables (see [13] and references therein). 
Our result is a generalization of this statement to the KP theory.
Note that the leaves of the foliation corresponding to non-zero values
of the periods of $\d E$ define a deformation of the space of hyperelliptic curves
in the moduli space of arbitrary curves with one puncture. 
\bigskip
\noindent{\it 4) The symplectic form for the Boussinesq Equation}

The Boussinesq equation corresponds to the case $n=3$, in
which
the operator $L$ is the third order
differential operator
$$
L=\p_x^3+u\p_x+v 
$$
The fundamental formula (29) becomes
$$
\omega_{\cal N}=-{1\over 3}\left(\delta u \wedge \int_{x_0}^x \delta v \,\d x +
\delta v \wedge \int_{x_0}^x \delta u\, \d x \right)
$$
which is a symplectic form on a space of two quasi-periodic functions 
$u=u(x,y),\ v=v(x,y)$ satisfying the constraints
$$
<u>_x=const,\ <v>_x= const. 
$$
The corresponding Poisson brackets have the form
$$
\{F,G\}={2\over 3} ({\delta F\over \delta u}\p_x{\delta G\over \delta v}+
{\delta G\over \delta u}\p_x{\delta F\over \delta v})\eqno(42)
$$
The leaf of the foliation defined by the differentials $\d p$ and $\d E=\d\Omega_3$
with the periods of $dE$ all zero, corresponds to a reduction of the KP
equation to the Boussinesq equation:
$$
u_t=2v_x-u_{xx},\ \
v_t=v_{xx}-{2\over 3}u_{xxx}-{2\over 3}uu_x\eqno(43) 
$$
Note that the usual form of the Boussinesq equation
$$
u_{tt}+({4\over 3}uu_x+{1\over 3}u_{xxx})_x=0 
$$
as an equation on one unknown function $u$, is the result of eliminating $v$ from
the system (43). 

The system (43) is a Hamiltonian system with the bracket (42) 
and the following Hamiltonian
$$
H={3\over 2}<v^2+uv_x+{1\over 3}u_x^2-{1\over 9}u^3>\eqno(44) 
$$
In terms of $H$ the equations for $u$ and $v$ assume the remarkably simple form
$$u_t={2\over 3}\p_x {\delta H\over \delta v},\ \ 
v_t={2\over 3}\p_x {\delta H\over \delta u}.\eqno(45) 
$$
We observe that the present Poisson bracket
seems to differ from the well-known Gelfand-Dickey bracket [14]
as well as from the generalized Gardner-Faddeev-Zakharov bracket.
It would be interesting to understand better the relation
between all of them. 
\bigskip
\centerline{\bf IV. OTHER INTEGRABLE MODELS}
\bigskip
It is straightforward to extend the preceding formalism
to more general situations, including matrix cases, higher symplectic forms,
and Calogero-Moser models. In this section, we discuss a few
examples.
\bigskip
\noindent {\it 1. The 2D Toda Lattice}

The 2D Toda lattice is the system of equations for the unknown functions
$\varphi_n=\varphi_n(t_+,t_-)$
$$
{\p^2\over \p t_+\p t_-}\varphi_n=e^{\varphi_n-\varphi_{n-1}}-
e^{\varphi_{n+1}-\varphi_{n}}. 
$$
It is equivalent to the compatibility conditions for the 
following auxiliary
linear problems
$$
\p_+\psi_n=\psi_{n+1}+v_n \psi_n,\ v_n=\p_+ \varphi_n, \ \
\p_-\psi_n=c_n\psi_{n-1},\ c_n=e^{\varphi_n-\varphi_{n-1}}. 
$$
A construction of  algebraic-geometric solutions 
to the hierarchy of this system was proposed in [15]. 

Let $\Gamma$ be a smooth genus $g$ algebraic curve with fixed local coordinates
$z_{\pm}$ in the neighborhoods of two punctures $P_{\pm}$. Let
$t=(t_{m,\pm})_{m=1}^{\infty}$ be the set of time parameters.
For any set
of $g$ points $\G_s$ in general position, the Baker-Akhiezer
function is the unique function
$\psi_n$  which is meromorphic on $\Gamma$ outside the punctures, has at most 
simple poles at the points $\G_s$,
and admits the following singularities near $P_{\pm}$
$$
\psi_n(t,z_{\pm})=z_{\pm}^{-\pm n}
\left(\sum_{s=0}^{\infty}\xi_s^{\pm}(n,t)z_{\pm}^{s}\right)
\exp(\sum_{m=1}^{\infty}t_{m,\pm}z_{\pm}^{-m}) ,\ \xi_0^+\equiv 1. 
$$
The times $t_{\pm}$ appearing in the Toda lattice correspond
to $t_{1,\pm}$ in the Baker-Akhiezer function. We obtain a solution
of the Toda lattice by setting
$$
\varphi_n={\rm log}\, \xi_0^-(n,t). 
$$
The integer $n$ acts as a discrete space variable.
It couples to the {\it quasi-momentum} $p$, which is the Abelian integral
characterized by the fact that $\d p$ has simple poles
at the punctures $P_{\pm}$ with residues $\mp 1$
respectively, and is real-normalized.
The differential which couples to the variable $t_+$
is the diffferential 
$\d\Omega_{+}$ which has a pole at $P_+$ of the form
$$
\d\Omega_+=d(z_+^{-1}+O(z_+)) 
$$

\medskip
\noindent{\bf Theorem 3}.
{\it The symplectic form $\omega_{\cal M}$ corresponding to the 
differentials $\d p$ and $\d E=\d\Omega_+$ is equal to
$$
\omega_{\cal M}=-\sum_{\alpha=\pm} {\rm Res}_{P_{\alpha}}
{<\delta \psi_n^+\wedge\delta L \psi_n>\over <\psi_n^+\psi>}\d p
=<\delta \varphi_n \wedge \delta v_n>\eqno(46)
$$
where $\varphi_n$ is the corresponding algebraic-geometric solution to the
2D Toda lattice, and $v_n=\p_+ \varphi$.}
\bigskip

\noindent{\it 2) Matrix Equations}

We show now that the algebraic-geometric symplectic
form $\omega_{\cal M}$ generated by two 
real-normalized differentials $\d p$ and $\d E$
having poles of orders $2$ and $n$ respectively at $N$ punctures 
$P_{\alpha}$, is a restriction of a symplectic form defined on
a space of linear operators with matrix $(N\times N)$ coefficients. 

The algebraic-geometric solutions to equations which have the zero-curvature
representation
$$
[\p_y-L, \p_t -A]=0\leftrightarrow L_y-A_y+[L,A]=0\eqno(47) 
$$
with
$$
L=\sum_{i=1}^n u_i(x,y,t)\p_x^i, \ A=\sum_{i=1}^m v_i(x,y,t)\p_x^i\eqno(48) 
$$
linear operators with matrix $(N\times N)$ coefficients,
was constructed in [11] (an updated version is in [8]). The construction is based
on the concept of vector Baker-Akhiezer function $\psi(x,y,t,z)$.
 
We fix a set of constants $a_{\beta}$ and a set of polynomials in $z^{-1}$
$$
q_{\beta}(z^{-1})=\sum_{j=1}^{m_{\beta}}q_{j,\beta}z^{-j}. 
$$
Then for any smooth genus $g$ 
algebraic curve $\Gamma$ with fixed local coordinates
$z_{\alpha}$ in the neighborhood of the punctures $P_{\alpha}$, and for
any set of $g+N-1$ points $\G_s\in \Gamma$ in general position, 
$\psi_{\alpha}$ is the unique function which 
is meromorphic outside the punctures, has at 
most simple poles at the points $\G_s$, and
has the following form in the neighborhood of the puncture $P_{\beta}$
$$
\psi_{\alpha}(x,y,t;z)=(\delta_{\alpha,\beta}+
\sum_{s=1}^{\infty}\xi_s^{\alpha,\beta}(x,y,t)z^{s}
)e^{z^{-1}x+a_{\beta}z{-n}y+q_{\beta}(z^{-1})t},
\ z=z_{\beta}.\eqno(49) 
$$

The vector Baker-Akhiezer function $\psi(x,y,t,z)$ 
is defined by $(\psi_{\alpha}(x,y,t,z))_{\alpha=1}^N$.
There exists then unique operators $L$ and $A$ of the form (48)
such that
$$
(\p_y-L)\psi(x,y,t,P)=0,\  (\p_t-A)\psi(x,y,t,P)=0.\eqno(50) 
$$
The coefficients of the operators can be derived from the 
following system of linear 
equations
$$
\sum_{i=0}^n u_{i}\sum_{l=0}^i C_i^l(\p_x^{i-l}\xi_{s+l})=\xi_{s+n}A,
\ A^{\alpha,\beta}=a_{\alpha}\delta^{\alpha,\beta},
\ s=-n,\ldots,-1,0,\eqno(51) 
$$
$$
\sum_{i=0}^m v_{i}\sum_{l=0}^i C_i^l(\p_x^{i-l}\xi_{s+l})=\sum_{j=1}^m
\xi_{s+j}\hat q_j,
\ \hat q_j^{\alpha,\beta}=q_{j,\alpha}\delta^{\alpha,\beta},\ 
s=-m,\cdots,-1,0,\eqno(52) 
$$

In particular, we have
$$
u_n=A,\ u_{n-1}=[\xi_1,A],\ u_{n-2}=[\xi_2,A]-n\p_x \xi_1,\ \ldots. 
\eqno(53)
$$
The defining condition (50) implies that the operators $L$ and $A$
satisfy the compatibility conditions (47), and hence
are solutions of the corresponding non-linear equations.

We also require the concept of the dual vector
Baker-Akhiezer function $\psi^+(x,y,t,P)$, which should be considered as
{\it row-vector}, whose components $\psi^+_{\alpha}$ are 
characterized by the following analytical properties.
Let $\d\Omega$ be a unique differential with poles of the form
$$
\d\Omega=\d(z_{\alpha}^{-1}+O(z_{\alpha})), 
$$
at the punctures $P_{\alpha}$, and with zeros at the marked points $\G_s$. 
In addition to $\G_s$, it has
$g+N-1$ other zeros which we denote $\G_s^+$. Then $\psi_{\alpha}^+$ 
is the unique function which 
is meromorphic outside the punctures,
has at most simple poles at the points $\G_s^+$, and
has the form
$$
\psi_{\alpha}^+(x,y,t,z)=(\delta_{\alpha,\beta}+
\sum_{s=1}^{\infty}\xi_{s}^{\alpha,\beta;+}(x,y,t)z^{s} )
e^{-z^{-1}x-a_{\beta}z^{-n}y-q_{\beta}(z^{-1})t},
\ z=z_{\beta}. 
$$
in the neighborhood of the puncture $P_{\beta}$. 
\bigskip
\noindent{\bf Theorem 4}. {\it  The restriction 
of the form
$$
\omega_{\cal M}=\delta \big(\ \sum_{s=1}^{g+N-1} p(\G_s)\,\d E(\G_s)
\big), \eqno(54)
$$
to the fibration ${\cal N}^{g+N-1}$
over the leaf $\M$ in $\M_g(n,2)$
given by real-normalized meromorphic differentials
$\d p$ and $\d E$ with
poles at the punctures of the form
$$
\d p=\d (z^{-1}+O(z)),\ \d E=\d( a_{\alpha}z^{-n}+O(z)), 
$$
is equal to}
$$
\omega_{\cal M}=-\sum_{\alpha=1}^{N}{\rm Res}_{P_{\alpha}}
{<\delta \psi^+\wedge \delta L\psi>\over
<\psi^+\psi>}\d p.\eqno(55) 
$$
\medskip
The proof of the theorem is essentially identical to that of the scalar case.
We observe that, although the proof relies on the
vector Baker-Akhiezer function as a function
which depends on a infinite number of times $t_{\alpha,i}$,  
the statement of the theorem itself uses only three marked times $(x,y,t)$.
\bigskip
As a concrete example, we discuss the {\it N-wave equation.} 
Consider 
the case $n=1$.  In this case the operator $L$ has the form
$$
L=A\p_x+u(x,y)
$$
where $A$ is the $N\times N$ matrix
$$
A^{\alpha,\beta}=a_{\alpha}\delta_{\alpha,\beta}
$$
and $u(x,y)$ is an $N\times N$ matrix with zero
diagonal entries
$ 
u^{\alpha,\alpha}=0$. 
As shown in [11] and references therein, for $m=1$,
the leaves corresponding to zero periods of $\d E$ 
give solutions to the so called $N$-wave equation.

The equations (51-52) imply that
$$
u^{\alpha,\beta}=(a_{\beta}-a_{\alpha})\xi_{1,\alpha,\beta}. 
$$
From the definition of the dual Baker-Akhiezer function it follows that
$$
\xi_1^{\alpha,\beta}+\xi_1^{\beta,\alpha;+}=0. 
$$
Thus the right hand side of the formula (55) defines 
the following symplectic 
form $\omega$ on the space of matrix-functions $u$ with zero diagonal elements
$$
\omega = \sum_{\alpha\neq \beta}{1\over a_{\alpha}-a_{\beta}}
\delta u^{\beta,\alpha}\wedge \delta u^{\alpha,\beta}.\eqno(56)
$$
\bigskip
\noindent{\it 3) Higher symplectic forms}

So far we have considered only cases where the 
differential $\d Q$ has poles of order $2$ at the punctures $P_{\alpha}$, or 
simple poles with resides $\mp 1$ at the punctures $P_{\alpha}^{\pm}$.
In these cases $\d Q$ may be identified with the
differential of the quasi-momentum
coupled with the marked variable $x$ in the case of differential operators, or
the discrete variable
$n$ in the case of difference operators.

We consider now the general case. Let $\d E$ and $\d Q$ be 
real-normalized differentials having at the puncture $P_{\alpha}$
the form
$$ 
\d E=a_{\alpha} d(z^{-n}+O(z)),\ \ \d Q=\d(q_{\alpha}(z^{-1})+O(z)) , 
$$
If the polynomials $q_{\alpha}(z^{-1})$ are the same as in (49),
the differentials $\d E$ and $\d Q$ reduce to the
differentials of the quasi-momenta for the corresponding
Baker-Akhiezer function with respect to the
variables $y$ and $t$.
\bigskip
\noindent{\bf Theorem 5}. {\it  The restriction of the form
$$
\omega_{\cal M}=\delta\big(\sum_{s=1}^{g+N-1} Q(\G_s)\d E(\G_s)\big)
\eqno(57)
$$
to a leaf of the corresponding foliation is equal to
$$
\omega_{\cal M}=\sum_{\alpha=1}^{N}{\rm Res}_{P_{\alpha}}
{<\delta \psi^+\wedge (L^{(1)}\delta A-A^{(1)}\delta L)\psi>\over
<\psi^+\psi>}\d p\eqno(58) 
$$
where $L^{(1)}$ and $A^{(1)}$ are the first descendants of the operators
$L$ and $A$ given in (50)}
$$
L^{(1)}=\sum_{i=1}^n iu_i(x,y,t)\p_x^{i-1}, \ 
A^{(1)}=\sum_{i=1}^m iv_i(x,y,t)\p_x^{i-1}.\eqno(59)
$$
\medskip
The proof of this theorem follows the
same lines as the proof of Theorem 2, with the
identity (31) replaced by the following generalization, whose
proof was also given in [8] (the additional
terms in [8] cancel on the leaves we are considering)
$$
\delta Q(E)={<\psi^+(L^{(1)}\delta A-A^{(1)}\delta L)\psi>\over
<\psi^+\psi>}{\d Q\over \d E}. 
$$

As an example, consider the case where the number of punctures $N$ is 1,
$n$ is $3$, and $m$ is $2$. The corresponding operators $L$ and $A$ are
$$
L=\p_x^3+u\p_x+v, \ A=\p_x^2+{2\over 3}u\eqno(60)  
$$
The operators $L^{(1)}$ and $A^{(1)}$ are then given by
$$
L^{(1)}=3\p_x^2+u, \ A=2\p_x^1  
$$
and the identity (58) becomes
$$
\omega=<{3\over 4} u \delta u \wedge\int_{x_0}^x\delta u\,\d x+
2 \delta v \wedge \int_{x_0}^x \delta v\,\d x +2 \delta u\wedge\delta v -
{3\over 2}\delta u_x \wedge \delta u > \eqno(61)
$$ 
The right hand side is well-defined on a space of two quasi-periodic 
functions $u(x,y)$ 
and $v(x,y)$ satisfying the constraints
$$
<u>_x=const, \ <v>_x=const,\ <u^2>_x=const.\eqno(62) 
$$
\bigskip
\noindent{\it 4) The elliptic Calogero-Moser System}

Until now we have only discussed systems
where the Lax pair $L,A$ does not contain a spectral parameter.
However, there is strong evidence that the present approach
is quite general, and can extend to this case as well.
A good example is the elliptic Calogero-Moser system,
which we discuss next.
\medskip
The elliptic Calogero-Moser system [16] is a system of $N$
identical particles on a line interacting with each other via the potential
$V(x)=\wp(x)$, where $\wp(x)=\wp (x|\omega,\omega')$ is the Weierstrass
elliptic function with periods $2\omega,\ 2\omega'$. The complete solution
of the elliptic Calogero-Moser system was constructed by algebraic-geometric
methods in [17]. There it was found that the equations of motion
$$
\ddot{x}_i=4\sum_{j\neq i} \wp'(x_i-x_j), \ \
\wp(x)={d\wp(x)\over dx}, \eqno(63)
$$
have a Lax representation {\it depending} on a spectral parameter $z$.
The Lax operator $L$ has the form:
$$
L_{ij}(t,z)=p_i\delta_{ij}+2(1-\delta_{ij})\Phi(x_i-x_j,z),
\ p_i=\dot x_i,
\eqno(64)
$$
with $\Phi(x,z)$ given by
$$
\Phi(x,z)={\sigma(z-x)\over \sigma(z)\sigma(x)}e^{\zeta(z)x}.
$$
In view of the Lax equation $\dot L=[M,L]$, the characteristic polynomial
$$
R(k,z)=\det (2k+L(t,z))
$$
is time--independent, and defines a time-independent spectral curve 
$\Gamma$ by 
the
equation
$$
R(k,z)\equiv\sum_{i=0}^N r_i(z)k^i=0 \eqno(65)
$$
where the $r_i(z)$ are elliptic functions of $z$. The Jacobians $J(\Gamma)$ of 
the
spectral curves $\G$ are levels of the involutive integrals of the system. 
In particular, we obtain angle variables $\varphi_i$ by choosing
$2\pi$-periodic coordinates on them.
It should be mentioned that although the exact solution to the
Calogero-Moser system in terms of $\Theta$-functions was found in [15], 
the
{\it action-variables} $a_i$ canonically conjugated to the 
{\it angle-variables}
$\varphi_i$ were found only relatively recently in [18]. There it was shown
that the Calogero-Moser sytem can be obtained through a Hamiltonian
reduction from the Hitchin system, and as a result, 
the action-variables $a_i$ are the periods of the differential
$kdz$ along $A$-cycles of the spectral curve $\Gamma$.

We show now that this statement can be obtained directly from our
approach. Let $C(P)=(c_1,\ldots,c_N)$ and $C^+(P)=(c_1^+,\ldots,c_N^+)$
be solutions of the linear equations
$$
(2k+L(z))C=0,\ C^+(2k+L(z))=0, \eqno(66)
$$
where $C$ is considered as a column-vector, $C^+$ as a row-vector, and
$P=(k,z)$ is a point of the spectral curve $\Gamma$.
If we normalize the eigenvectors $C$ and $C^+$ by the conditions
$$
\sum_{i}c_i\Phi(-x_i,z)=1,\ \sum_{i=1}c_i^+\Phi(x_i,z)=1, \eqno(67)
$$
then it follows from [17] that $C(P)$ and $C^+(P)$ are meromorphic
functions on $\Gamma$ outside the points $P_{\alpha}$ on $\G$ corresponding to
$z=0$, and have $N$ poles. We denote these poles by $\G_1,\cdots,\G_N$
and $\G_1^+,\ldots,\G_N^+$, respectively.
Near the points $P_{\alpha}$, the components of these vectors have the form
$$
c_i(P)=(c_i^{\alpha}+O(z))e^{x_iz^{-1}}, \
c_i^+(P)=(c_i^{\alpha,+}+O(z))e^{-x_iz^{-1}}\eqno(68)
$$
and
$$
\eqalignno{c_i^1&=1; \sum_i c_i^{\alpha}=0 \ \ {\rm for} \ \alpha>1 \cr
c_i^{1,+}&=1; \sum_i c_i^{\alpha,+}=0 \ \ {\rm for} \ \alpha>1&(69)\cr}
$$
The formalism in this paper applies exactly as before
to produce the following formula for the symplectic form for
the Calogero-Moser system
$$
2\delta(\sum_{s=1}^N k(\G_s)dz) =
\sum_{\alpha=1}^N{\rm Res}_{P_{\alpha}}{<\delta C^+\wedge \delta L C>
\over <C^+C>}dz, \eqno(70)
$$
with $<f^+ g>$ the usual pairing between vectors and co-vectors.
Substituting in the expansion (68-69), we obtain at once
$$
\delta(\sum_{s=1}^N k(\G_s)dz) ={1\over 2} \sum _i^N \delta p_i\wedge \delta x_i.
\eqno(71)
$$
\bigskip

\centerline{\bf V. N=2 SUPERSYMMETRIC SU($N_c$) GAUGE THEORIES}
\bigskip
We turn now to a discussion of
the universal configuration space $\M_g(n,m)$
and of the symplectic forms
$\omega_{\cal M}$ in the context of
N=2 supersymmetric Yang-Mills theories with gauge group SU($N_c$).
\medskip
We consider gauge theories with $N_c$ colors and
$N_f$ flavors. The field content is an N=2 chiral multiplet and $N_f$
hypermultiplets of bare masses $m_i$. The N=2 chiral
multiplet contains a complex scalar field $\phi$ in the adjoint
representation of SU($N_c$). The flat directions in the potential correspond to
$$
<\phi>=\sum_{i=1}^{N_c-1}a_iH_i
$$ 
with $H_i$ a basis for the Cartan subalgebra.   
Thus the theory admits a $N_c-1$ dimensional space of vacua,
which can be parametrized by the order parameters
$$
u_k\equiv Tr<\phi^k>,\ k=2,\cdots,N_c\eqno(72)
$$
It is often more convenient to work with the parameters $s_i$,
$i=0,\cdots,n-2$, defined recursively by $s_0=0$, $s_1=u_1=0$, and
$$
ks_k+\sum_{i=1}^ks_{k-i}u_i=0\eqno(73)
$$
In the weak coupling limit, the $s_k$'s correspond to
$k-th$ symmetric polynomials
in $\phi$, but they do not of course
admit in general such a simple interpretation.
In the N=1 formalism, the effective Lagrangian is of the
form
$$
{\cal L}={\rm Im}{1\over 4\pi}\big[\int d^4\theta\partial_i{\cal F}(A)\bar A^i
+
{1\over 2}\int d^2\theta\partial_i\partial_j{\cal F}(A)W^iW^j\big]
$$
where $A_i$ is an N=1 chiral superfield with scalar component
$a_i$. The holomorphic pre-potential ${\cal F}$ as well as the spectrum of BPS
states can all be determined by finding a fibration of spectral curves
as well as of Seiberg-Witten forms $\d\lambda$ over the
moduli space of vacua
$$
\eqalignno{&a\equiv\oint_A\d \lambda,\ a_D\equiv\oint_B\d\lambda,
\ {\partial{\cal F}\over\partial a_i}=a_{D,i}\cr
&M_{BPS}^2=2|Z|^2,\ Z=\sum_{i=1}^{N_c}[n_e^ia_i+n_m^ia_{D,i}]+\sum_{i=1}^{N_f}S_i
m_i
&(74)\cr}
$$
Here the $S_i$'s are U(1) charges corresponding to global symmetries, and $n_e^i$
and $n_m^i$ are respectively electric and magnetic charges. Some key requirements
for $\Gamma$ and $\d\lambda$ are e.g. consistency with
the weak coupling evaluation of the monodromies
of the effective Lagrangian as the gauge
bosons become massless, with instanton contributions,
with the
classical limit, and with an infinite mass limit
reducing the number of flavors [1][3]. 
\bigskip
We shall construct the desired fibration of spectral curves
and vector bundle simply
by identifying the moduli space
of gauge vacua with an appropriate leaf ${\cal M}$
in our universal configuration space $\M_g(n,m)$,
for a suitable choice of $n,m $ and $g$.
Thus set, in the notation of Section II,
$$
\eqalignno{
g&=N_c-1, (P_{\alpha})=(P_+,P_-;P_1,\cdots P_{N_f})\cr
m&=(m_+,m_-;m_1,\cdots,m_{N_f})=(1,1;0,\cdots,0)\cr
n&=(n_+,n_-;n_1,\cdots,n_{N_f})=(0,0;0.\cdots,0)&(75)\cr}
$$
In the resulting configuration space $\M_g(n,m)$, consider
the leaf $\M_{N_c,N_f}$ characterized by the following conditions on the singular
parts of the Abelian integrals $E$ and $Q$
$$
\eqalignno{
R_{\alpha}^E&=1\ {\rm for}\ 1\leq \alpha\leq N_f,\ R_-^E=N_c-N_f\cr
R_{\alpha}^Q&=0\ {\rm for}\ 1\leq \alpha\leq N_f,\ R_-^Q=0\cr
T_{\alpha,0}&=-m_i,\ 1\leq i\leq N_f,\ T_{+,0}=0\cr
T_{+,k}&=N_c2^{-1/N_c},\ k=1\cr
T_{-,k}&=(N_c-N_f)({\Lambda\over2})^{1/(N_c-N_f)}&(76)\cr}
$$
and the following conditions on their periods
$$
\oint_{A}\d Q=\oint_{B}\d Q=0\eqno(77)
$$
$$
{1\over 2\pi i}\oint_{A}\d E=m\in{\bf Z}^{N_c-1},{1\over 2\pi i}\oint_{B}\d E= n
\in{\bf Z}^{N_c-1}\eqno(78)
$$
The conditions on $Q$ imply that $Q$ is actually
a well-defined meromorphic {\it function} on $\Gamma$,
with only simple poles at $P_{\pm}$.
The third condition in (76)
means that $Q(P_i)=-m_i$. Now a point of the
universal configuration space also requires
a jet of coordinates at each puncture $P_{\pm}$ and $P_i$.
In the present case, these jets are provided by
the function $Q$, more precisely, by $Q^{-1}$
at $P_{\pm}$, and by $Q+m_i$ at $P_i$. 

The form $\d E$ is a meromorphic form with only simple poles
at $P_i$ and $P_{\pm}$. Its residues at $P_i$ are $1$,
while its residues at $P_+$ and $P_-$ are respectively
$-N_c$ and $N_c-N_f$. The form $\d\lambda=Q\d E$ is a single-valued meromorphic
1-form on $\Gamma$. We identify it with the Seiberg-Witten
form of the gauge theory. Since its residues at $P_i$, $1\leq i\leq N_f$,
are $-m_i$, we recognize the parameters $m_i$, $1\leq i\leq N_f$, as the masses
of the hypermultiplets in the gauge theory. The residues of $\d\lambda$
depend then linearly on the masses, as they should in order for
the BPS spectrum (c.f. (74)) to be expressible in terms of the periods of
$\d\lambda$. We shall see shortly
that the remaining parameter $\Lambda$ in (76), can be
interpreted as essentially the dynamically generated scale
of the gauge theory.     
\medskip
We pause briefly to discuss the holomorphicity
of the derivatives along $\M_{N_c,N_f}$ of $Q\d E$.
As we have seen in Section II, the notion of derivatives
requires a connection, which is in the case provided by
by the level sets of the Abelian integral $E$. It is with
respect to this connection, $\nabla^E$, that
the derivatives of $Q\d E$ are holomorphic. However, the roles
of $E$ and $Q$ are almost interchangeable, and we can consider
$\nabla^Q(Q\d E)$ as well. On functions $f$, the two connections
are related by
$$
\nabla_X^Ef=\nabla_X^Qf-{\d f\over\d E}\nabla_X^QE\eqno(79)
$$
With respect to the connection $\nabla^Q$, the derivatives
of $Q\d E$ will develop poles. However, interchanging
the roles of $E$ and $Q$, we can see that $\nabla^Q(E\d Q)$
is holomorphic. In fact, the equation (79) implies
that
$$
\nabla^E(Q\d E)=(\nabla^EQ)\d E=
(\nabla^QQ-{\d Q\over\d E}\nabla^QE)\d E=-\nabla^Q(E\d Q)
\eqno(80)
$$
This shows that at the level of derivatives along
$\M_{N_c,N_f}$, the two forms $Q\d E$ and $E\d Q$
are interchangeable. However, for the leaf $\M_{N_c,N_f}$
we are discussing, only $Q\d E$ is well-defined.
\medskip

We derive now the equation of the curve $\Gamma$.
First, it follows from the existence of a meromorphic function
$Q$ with exactly two simple poles that $\Gamma$
is a hyperelliptic Riemann surface. Next,
the integrality conditions (78) imply that, although $E$
is a multiple-valued Abelian integral, the function
$$
w={\rm exp}\,E\eqno(81)
$$
is a well-defined meromorphic function on $\Gamma$, with only poles
at $P_{\pm}$, of order respectively $N_c$ and $N_f-N_c$. 
As a consequence, there exists
polynomials $A(Q)$ and $B(Q)$ of degrees respectively $N_c$ and
$N_f$ such that
$$
w+{B(Q)\over w}=2A(Q)\eqno(82)
$$ 
At the (finite) zeroes $P_i$ of $w$, $Q(P_i)=-m_i$, and thus
$B(Q)$ must be of the form
$$
B(Q)=\tilde\Lambda\prod_{i=1}^{N_f}(Q+m_i)\eqno(83)
$$
for a suitable constant $\tilde\Lambda$. We shall show
that this parameter $\tilde\Lambda$ is the same as the
earlier $\Lambda$ parametrizing the leaf. For this, we consider
the expansion of the Abelian integral $p$ near $P_-$ in terms of $Q$.
Recall that the coordinate $z$ appearing in the definition
of $T_{-,k}^Q$ is defined by $p=-N_c{\rm log}\,z$. Thus the
fourth condition in (76) means that
$Q=2^{-1/N_c}z^{-1}+O(1)$, and we have near $P_-$
$$
E=N_c{\rm log}\,Q+{\rm log}\,2+O(Q^{-1})\eqno(84)
$$  
Similarly, near $P_+$, we have 
$$
E=(N_c-N_f){\rm log}\,Q+{\rm log}\,2+{\rm log}{\Lambda\over 4}+O(Q^{-1})\eqno
(85)
$$
This means in particular that the parameter $\Lambda$
of (76) can be interpreted as the regularized version of the
following logarithmically divergent integral
$$
{\rm log}\,{\Lambda\over 4}=\big(\int_{P_+}^{P_-}\d E\big)
+(2N_c-N_f){\rm log}\,Q\vert_{Q\rightarrow\infty}\eqno(86)
$$
This is a confirmation of the interpretation of $\Lambda$
as a scale, since the classical limit $\Lambda\rightarrow0$
does correspond to the two poles $P_{\pm}$ becoming infinitely
separated, and to the curve $\Gamma$ degenerating to the classical
curve. We can now determine $\tilde\Lambda$ by evaluating
the right hand side of (86), using the equation for $w$. We find
$$
\int_{P_+}^{P_-}\d E
=-{\rm log}{A+\sqrt{A^2-B}\over A-\sqrt{A^2-B}}_{\vert_Q=\infty}
=-(2N_c-N_f){\rm log}\, Q+{\rm log}\,{\tilde\Lambda\over 4}
$$
so that $\Lambda$ agrees with $\tilde\Lambda$.
\medskip
We can also write the form $\d\lambda=Q\d E$ in terms of the
polynomials $A(Q)$ and $B(Q)$. Setting $y=w-A(Q)$, we find
$$
\d\lambda
={Q\over y}(A'-{1\over 2}{B'\over B}(A-y))\d Q
\eqno(87)
$$
The poles $P_1,\cdots,P_{N_f}$ of this form $\d\lambda$ occur
on only one sheet of the hyperelliptic surface $\Gamma$.
The symmetric version with poles on both sheets
adopted by Seiberg and Witten [2] (see also
Hanany and Oz [3]) can be obtained by
symmetrizing
$$
\d\tilde\lambda={Q\over y}(A'-{1\over 2}{AB'\over B})\d Q\eqno(88)
$$
We note that this does not affect the physics,
since the difference ${QB'\over 2B}\d Q$ is $y$-independent,
and hence has zero periods.
\medskip
As we can see from Theorem 1, a natural parametrization of the leaf
${\cal M}_{N_c,N_f}$ is in terms of the periods $a_i$
of $\d\lambda=Q\d E$ along $A$-cycles. This
is particularly attractive from the viewpoint of gauge
theories, since the $a_i$'s are the
scalar components of the light superfields
in the effective theory. The dual fields $a_{D,i}$ as well
as the effective Lagrangian ${\cal F}$ are already
determined by our formalism without having to
parametrize ${\cal M}_{N_c,N_f}$ specifically in
terms of the order parameters $u_k$ of the gauge theory.
The same is true of all the information we have obtained so far
on the equation of the curve $\Gamma$ itself.

To write $\Gamma$ in terms of the $u_k$'s, we
consider the expansion of $E$ near $P_+$, and set
$$
E=N_c{\rm log}\,Q+{\rm log}\,2+{\rm log}\,(1+\sum_{i=1}^{\infty}s_iQ^{-i})
\eqno(89)
$$
In view of the third constraint in (76), the coefficient $s_1$
is 0. The next $N_c-1$ coefficients $s_2,\cdots,s_{N_c}$
provide independent coordinates for the leaf ${\cal M}_{N_c,N_f}$.
We identify them with the order parameters $s_i$ of
the gauge theory, as defined by (73).    

With this identification, it is easy to determine all the coefficients of $A(Q)$
in terms of the $s_k$'s,
since the Abelian integral $E$ is also given by
the equation
$$
E={\rm log}\,w={\rm log}\,(A+(A^2-B)^{1/2})\eqno(90)
$$
We shall work it out explicitly when $N_f\leq N_c+1$. Set
$A(Q)=\sum_{i=0}^{N_c}c_{i}Q^{N_c-i}$. In this case, we may rewite 
on one hand the expansion (89) as
$$
E={\rm log}\,2+{\rm log}\,(Q^{N_c}+\sum_{i=2}^{N_c}s_iQ^{N_c-i}+O(Q^{-1}))\eqno(91)
$$
and on the other hand, the expansion (90) as
$$
E={\rm log}\,2+{\rm log}\,(A-{1\over 4}{B\over A}+O(Q^{-1}))\eqno(92)
$$
Comparing the two equations, we find that $A(Q)$ is monic, and that
$c_1=s_1=0$. This implies in turn that
$A^{-1}=Q^{-N_c}(1+O(Q^{-2}))$, and we obtain
$$
Q^{N_c}+\sum_{i=2}^{N_c}s_iQ^{N_c-i}
=A-{1\over 4}B+O(Q^{-1})
$$
Altogether, setting 
$$y=w-A(Q),\ t_k(m)=\sum_{i_1<\cdots<i_k}m_{i_1}\cdots m_{i_k},
$$
and making the identification
$\Lambda\rightarrow \Lambda_{N_c,N_f}^{2N_c-N_f}$,
where $\Lambda_{N_c,N_f}^{2N_c-N_f}$ is the dynamically
generated scale of the theory, we have shown that
\medskip
\noindent{\bf Theorem 6}. {\it Let $\M_{N_c,N_f}$ be the leaf
of the universal configuration space $\M_g(n,m)$ defined
by the level sets (75-78), and define $\d\lambda$ as $\d\lambda=Q\d E$.
Then $\M_{N_c,N_f}$ can be parametrized by the $A$-periods
$a_i$'s of $\d\lambda$, or by the coefficients $s_2,\cdots,s_{N_c}$
of (73). The equation of the curves $\Gamma$ fibering
over $\M_{N_c,N_f}$ is $y^2=A(Q)^2-B(Q)$, with $B(Q)$ given by
(83). For $N_f\leq N_c+1$, the equation of the curve
$\Gamma$ takes the form suggested by Hanany and Oz}
$$
y^2=\big(\sum_{i=0}^{N_c}s_iQ^{N_c-i}+{
\Lambda_{N_c,N_f}^{2N_c-N_f}\over 4}\sum_{k=0}^{N_f-N_c}
t_k(m)Q^{N_f-N_c-k}\big)^2-
\Lambda_{N_c,N_f}^{2N_c-N_f}\prod_{i=1}^{N_f}(Q+m_i)
\eqno(93)
$$
\bigskip
Finally, we note a possible interpretation, in the context
of integrable models, for {\it both} the Abelian integrals
$E$ and $Q$ appearing in the form $\d\lambda$
of supersymmetric gauge theories. For
pure SU($N_c$), $Q$ corresponds to the eigenvalues
of the operator $(L\psi)_n=\psi_{n+1}+v_n\psi_n+c_{n-1}\psi_{n-1}$
defining the periodic Toda chain, and ${\rm exp}\,E$ to
the eigenvalues of the monodromy operator
$(T\psi)_n=\psi_{n+N_c}$. In presence of
hypermultiplets, the periodic boundary condition
gets shifted to an aperiodic Toda chain, with
the "monodromy" operator $T$ rather of the form
$(T\psi)_n=\psi_{n+N_c}+
\sum_{\kappa=1}^{N_f}\lambda_{n,\kappa}\psi_{n+N_c-\kappa}$.

In [18], Gorsky et al. suggest other
representations in terms of spin chains. It would
be interesting to understand better the
relation between these different constructions.
\bigskip
\centerline{\bf APPENDIX}
\bigskip
In this Appendix, we provide a proof of Theorem 1.
\medskip
Assume that the differentials of all the coordinates are linearly
dependent at a point of our universal configuration space. 
Then there exists a one
parameter deformation of this point (i.e. a one parameter
deformation of the curve $\Gamma^t$ and of all the other data) with
the derivative $\p_{t}|_{t=0}$ of all the data equal to
zero. In particular
$$
\p_{t}(Q(E,t)\d E)|_{t=0}=0, \eqno(A.1)
$$
since the left-hand side is a holomorphic
differential on $\Gamma^0$ with zero periods and must be
identically zero. Let $\G$ be a point of the zero divisor of $\d E$ on $\Gamma^0$.
For simplicity, we consider first the case where this point is a simple zero.
Then, locally, the equation $\d E(\gamma(t))=0),\ \G(t)\in \Gamma^{t}$ defines
a deformation of $\G$. In a neighborhood of our point on the universal
configuration space, the Abelian integral
$Q$ is a holomorphic function in this neighborhood, and has an expansion of
the form
$$
Q(E,t)=q_0(t)+\sum_{i=1}^\infty q_i(t)(E-E_{t})^{i/2}, \eqno(A.2)
$$
where $E_{t}$ is a critical value of Abelian integral $E$:
$$
E_{t}=E(\G_{t}) \eqno(A.3)
$$
and the $q_i$'s are analytic functions of the variable $t$. Since
$$
\d Q(E,t)=q_1(t){\d E\over \sqrt{E-E_{t}}}+q_2\d E+O(E-E_{t})^{1/2})\d E
\eqno(A.4)
$$
and thus $q_1(0)\neq 0$ in view of the assumptions
of Theorem 1. The equality
$$
\p_{t}Q(E,t)|_{t=0}={q_1(0)\p_{t}E_{t}|_{t=0}\over
\sqrt(E-E_{t})}+O(1), \eqno(A.5)
$$
and (A.2) imply then
$$
\p_{t}E_{t}|_{t=0}=0. \eqno(A.6)
$$
Next, we show that (A.6) implies in turn that
$$
\p_{t}\omega_i(E,t)|_{t=0}=0\eqno(A.7)
$$
where
$$
\omega_i(E,t)=\int_{P_1^t}^E\d\omega_i 
$$
is the Abelian integral of the normalized holomorphic differential $\d\omega_i$.
Indeed, 
the left hand side of (A.7) is an Abelian integral with possibly poles
at the zeros of $\d E$. Its expansion at these zeros similar to (A.2) shows
that it is holomorphic at these points due to (A.6). Thus the left
hand side of (A.7) is a holomorphic Abelian integral with zero $A$-periods,
and must be identically 0. Its $B$-periods must also be zero, and hence
$$
\p_{t}\tau_{ij} (t)|_{t=0}=0, 
$$
Here $\tau_{ij}(t)$ the period matrix of 
$\Gamma^t$. In view of the infinitesimal Torelli theorem, this can only be true
if $\Gamma^0$ is a hyperelliptic curve, and $\p_{t=0}$ is transversal to
the moduli space of hyperelliptic curves. In order to
finish the proof that up to the order $O(t^2)$ the curve $\G^t$ does not
change , we shall show that if (A.6) is fulfilled for some
hyperelliptic curve $\Gamma^0$, then the vector $\p_{t=0}$ is tangent to the moduli
space of hyperelliptic curves.

We fix on $\Gamma^0$ a pair of distinct points $P^{\pm}$
for which there exists a function $\lambda$ with simple poles at these
points and holomorphic everywhere else. We may choose them so that
$\d E(P^{\pm})\neq 0$. Let
$$
\lambda =a^{\pm}(E-E_{\pm}^0)+b^{\pm}+O(E-E_{\pm}^0),\
E_{\pm}^0=E(P^{\pm}) \eqno(A.8)
$$
be the expansions of $\lambda$ at the points $P^{\pm}$.
Then on $\Gamma^t$, there exists a unique meromorphic real-normalized
Abelian integral
$\lambda(E,t)$, with simple poles at the points $P^{\pm,t}$ given by
$E(P^{\pm,t})\equiv E_{\pm}^0)$, and whose expansion
near these points has the form of the right hand side of (A.8)
identically in $t$. Arguing just as before, we can conclude that
$$
\p_t \lambda(E,t)|_{t=0}=0. \eqno(A.9)
$$
This means that, up to order $O(t^2)$, the periods of $d\lambda(E,t)$
are the same as
the periods of $d\lambda(E,0)$, 
and thus equal to zero. The function $\lambda (t)$ is a single-valued function with
only two simple poles, the curve $\Gamma^t$
is hyperelliptic, and $\p_{t=0}$ is a tangent vector to the moduli space of
hyperelliptic curves.

This completes the proof that, up to order $O(t^2)$,
the curve $\Gamma^0=\Gamma^t$ does not change. Now we have to prove that the
punctures and the jets of local coordinates also do not change.

For $g=0$ or $1$, the nontrivial automorphism group of the curve
allows us to fix the puncture $P_1$. For $g>1$, there exists
a holomorphic differential $\d\omega_0$ with 
$$
\d\omega_0 (P_{\alpha(0)})\neq 0 \eqno(A.10)
$$
with $\d E$ not vanishing at least one of its zeros
$$
\d\omega_0(p_0)=0,\ \d E(p_0)\neq 0. \eqno(A.11)
$$
As before, as a consequence, the Abelian integral
$$
\omega_0(P) = \int_{P_1}^P \d\omega_0 \eqno(A.12)
$$
must satisfy
the relation
$$
\p_t\omega_0(E,t)|_{t=0}=0. \eqno(A.13)
$$
wnen it is considered as a function of $E$.
The derivatives with fixed $E$ or $P$ 
are related to each other by (c.f. (80))
$$
\p_t \omega_0(P)=\p_t \omega_0(E,t)+\p_tE(P,t){\d\omega_0\over \d E}\eqno(A.14)
$$
Taking the derivative of (A.12) at $P=p_0$,
we get
$$
0=\p_t\omega_0(p_0)|_{t=0}=-\left(\p_t z(P_1(t))|_{t=0}\right)
{\d\omega_0\over \d z}(P_1(0)), \eqno(A.15)
$$
where $z(P)$ is any local coordinate in a neighborhood of $P_1(0)$.
Thus, up to order $O(t^2)$,
the point $P_1(t)=P_1(0)$ does not change.

At the points $P_{\alpha}(t)$ the function $E$ has a constant value (which is
infinity). In view of (A.12) and the fact that $\p_t P_1|_{t=0}=0$,
we have then
$$
\big(\p_t \int_{P_1}^{P_{\alpha}(t)}\d\omega_0\big) |_{t=0}=0.
\eqno(A.16)
$$
Since $\d\omega_0$ does not vanish
at $P_{\alpha(0)}$ (c.f. (A.10)), we conclude that
$\p_t P_{\alpha}(t)|_{t=0}=0$. 

The next step is to prove that jets of local coordinates at the punctures
do not change. We proceed in the same way as before and consider the
derivatives of the real-normalized  differentials $\d\Omega_{(k)}$
which have poles of the form
$$
\d\Omega_k=\d(E^{k_{\alpha}/n_{\alpha}}+O(1)),
k=(k_1,\ldots, k_N), \ k_{\alpha}\leq n_{\alpha}, \eqno(A.17)
$$
at the punctures $P_{\alpha}$, and are holomorphic otherwise.
Exactly in the same way as before, our assumptions
imply that
$$
\p_t \Omega_k(E,t)|_{t=0}=0. \eqno(A.18)
$$
The last equalities imply that the $n_{\alpha}$-jets of local coordinates which
were
used to define $E$ do not change.
The theorem is proved under the assumption that all zeros of $\d E$ on
$\Gamma^0$ are
simple. 

We consider now the general case, where $\d E$ may have multiple zeroes.
Let $D=\sum \mu_s \gamma_s$ be the zero divisor of $\d E$. The degree
of this divisor
is equal to
$$
\sum_s \mu_s=2g-2+N+\sum_{\alpha=1}^N n_{\alpha}. \eqno(A.19)
$$
Consider a small neighborhood of $\gamma_s$, viewed as a point of the
fibration ${\cal N}$, above the original data point
in the universal configuration space. Viewed as a
function on the fibration, $E$ is a
deformation
of its value $E(P,m_0)$ above the original
data point, with multiple critical points $\gamma_s$.
Therefore, on each of the corresponding curve, there exists a local coordinate
$w_s$ such that
$$
E=w_s^{\mu_s+1}(P,m)+\sum_{i=0}^{\mu_s-1} E_{s,i}(m)w_s^i(P,m). \eqno(A.20)
$$
The coefficients $E_{s,i}(m)$ of the polynomial (A.20) are well-defined
functions on $\M_g(n,m)$. If $\mu_s=1$, then $E_{s,0}$ coinsides with
the critical value $E(\G)$ from (A.3). In the same way as before,
we can prove that (A.1) implies that
$$
\p_t E_{s,i}(t)|_{t=0}=0, \eqno(A.21)
$$
after which the arguments become identical to the previously
considered case.
Theorem 1 is proved.
\vfill\break
\centerline{\bf ACKNOWLEDGEMENTS}
\bigskip
The authors would like to thank
E. D'Hoker, A. Gorsky, A. Marshakov,
A. Mironov, and A. Morozov
for very helpful conversations.
\vfill\break

\centerline{\bf REFERENCES}
\bigskip
\item{[1]} N. Seiberg and E. Witten, "Electric-magnetic duality, monopole
condensation, and confinement
in N=2 supersymmetric Yang-Mills theory", Nucl. Phys. B 426 (1994) 19,
hepth/9407087;
"Monopoles, duality, and chiral
symmetry breaking in N=2 supersymmetric QCD", Nucl. Phys. B 431 (1994)
484, hepth/9408099.
\item{[2]} A. Klemm, W. Lerche, S. Yankielowicz, S. Theisen, 
"Simple singularities and N=2 supersymmetric Yang-Mills theory", Phys. Lett. 344 B
(1995) 169; \hfill\break
P.C. Argyres and A.E Faraggi, "The vacuum structure and spectrum
of N=2 supersymmetric SU(N) gauge theory", Phys. Rev. Lett. 73 (1995)
3931, hepth/9411057;\hfill\break
M.R. Douglas and S. Shenker,  "Dynamics of SU(N) supersymmetric gauge
theory", hepth/9503163;\hfill\break
P.C. Argyres and M.R. Douglas, "New phenomena in SU(3) supersymmetric
gauge theory", hepth/9505062.
\item{[3]} A. Hanany and Y. Oz, "On the quantum moduli space of vacua
of N=2 supersymmetric SU(N) gauge theories", hepth/9505075\hfill\break
P. Argyres, R. Plesser, and A. Shapere, "The Coulomb
phase of N=2 supersymmetric QCD", Phys. Rev. Lett. 75 (1995) 1699-1702,
hepth/9505100\hfill\break
J. Minahan and D. Nemeshansky, "Hyperelliptic curves for
supersymmetric Yang-Mills", hepth/9507032\hfill\break
P.C. Argyres and A. Shapere, "The vacuum structure
of N=2 super-QCD with classical gauge groups", hepth/9509175\hfill\break
A. Hanany, "On the quantum moduli space
of vacua of N=2 supersymmetric gauge theories", hepth/9509176.
\item{[4]} U.H. Danielsson and B. Sundborg, "Exceptional Equivalences
in N=2 Supersymmetric Yang-Mills Theory", USITP-95-12, UUITP-20/95\hfill\break
U.H. Danielsson and B. Sundborg, "The moduli space
and monodromies of N=2 supersymmetric SO(2r+1) Yang-Mills theory, 
hepth/9504102;
\hfill\break
A. Brandhuber and K. Landsteiner, "On the monodromies of N=2 supersymmetric
Yang-Mills theory with gauge group SO(2n)", hepth/9507008
\item{[5]} A. Gorsky, I. Krichever, A. Marshakov, A Mironov, and A.
Morozov, "Integrability and Seiberg-Witten exact solutions", 
Phys. Lett. B 355 (1995) 466, hepth/9505035;\hfill\break
E. Martinec and N. Warner, "Integrable systems and
supersymmetric gauge theory", hepth/9509161, hepth/9511052\hfill\break
A.Gorsky and A.Marshakov, ``Towards effective topological
theories on spectral \hfill\break
curves'', hepth/9510224 \hfill\break
T. Nakatsu and K. Takasaki, "Whitham-Toda hierarchy and
N=2 supersymmetric Yang-Mills theory", hepth/9509162\hfill\break
H. Itoyama and A. Morozov, "Prepotential and the
Seiberg-Witten theory", \hfill\break
hepth/9511126, hepth/9512161, hepth/9601168\hfill\break
A. Marshakov, ``Exact solutions to quantum field theories and integrable
equations'', hepth/9602005 
\item{[6]} R. Donagi and E. Witten, "Supersymmetric
Yang-Mills theory and integrable systems", hepth/950101;\hfill\break
E. Martinec, "Integrable structures in supersymmetric
gauge and string theory", hepth/9510204
\item{[7]} H. Flaschka, M.G. Forest, and D.W. McLaughlin,
"Multiphase averaging and the inverse spectral solution
of the KdV equation", Comm. Pure Appl. Math. 33 (1980) 739-784\hfill\break
B.A. Dubrovin and S.P. Novikov, "The Hamiltonian formalism
for 1D systems of hydrodynamic type
and the Bogolyubov-Whitham averaging method", Sov. Math. Doklady 27 (1983) 665-669;
\hfill\break
I.M. Krichever, "The dispersionless Lax equation and
topological minimal models", Comm. Math. Phys. 143 (1992) 415-429;\hfill\break
B.A. Dubrovin, "Hamiltonian formalism of Whitham hierarchies
and topological Lan\-dau-Ginzburg models", Comm. Math. Phys. 145 (1992) 
195-207. 
\item{[8]} I.M. Krichever, "The averaging method for 2D systems", 
Funct. Anal. Appl. 22 (1988) 200-212.
\item{[9]} I.M. Krichever, "The $\tau$-function
of the universal Whitham hierarchy, matrix models,
and topological field theories",
Comm. Pure Appl. Math. 47 (1994) 437-475.
\item{[10]} A. Gorsky, A. Marshakov, A. Mironov, and A. Morozov, 
"N=2 supersymmetric QCD and integrable
spin chains: rational case $N_f<2N_c$", hepth/9603140
\item{[11]} I.M. Krichever, "Methods of algebraic
geometry in the theory of non-linear equations", Russian Math. Surveys
32 (1977) 185-213.
\item{[12]} S.P. Novikov and A. Veselov, "On Poisson brackets
compatible with algebraic geometry
and Korteweg-de Vries dynamics on the space
of finite-zone potentials", Sov. Math. Doklady 26 (1982) 357-362;
"Poisson brackets and complex tori", Proc. Steklov Inst.
Math. 165 (1985) 53-65.
\item{[13]} S.P. Novikov, S.V. Manakov, L.P. Pitaevski, V.E. Zakharov,
"{\it Theory of Solitons}", Contemporary Soviet Mathematics,
Plenum Press (1984), New York'\hfill\break
H. Flaschka and D. McLaughlin, "Canonically conjugate
variables for KdV and the Toda lattice with periodic
boundary conditions", Prog. Ther. Phys. 55 (1976) 438-456.
\item{[14]} I.M. Gelfand and L.A. Dickey, "Fractional powers of
operators and Hamiltonian systems", Funct. Anal. Appl. 10 (1976) 13-29;\hfill\break
L.A. Dickey, "{\it Soliton equations and
Hamiltonian systems}", World Scientific (1991), Singapore.
\item{[15]} I.M. Krichever, "The periodic nonabelian Toda lattice
and its two-dimensional generalization", Uspekhi Mat. Nauk 36 (1981) 72-77.
\item{[16]} F. Calogero, "A method to generate soluble non-linear
evolution equations", Lett. Nuovo Cimento 14 (1975) 443-447\hfill\break
J. Moser, "Three integrable Hamiltonian systems connected
with isospectral deformations", Advances Math. 16 (1975) 197-220;\hfill\break
A.M. Perelomov, "Completely integrable
classical systems connected with semisimple
Lie algebras", Lett. Math. Phys. 1 (1977) 531-540.
\item{[17]} I.M. Krichever, "Elliptic solutions of the
Kadomtsev-Petviashvili equation and
integrable systems of particles", Funct. Anal. Appl. 14 (1980) 282-290.
\item{[18]} A. Gorsky and N. Nekrasov, "Elliptic Calogero-Moser system from
two-dimensional current algebra", hepth/9401021.

\end
 
\end